%% file: main.tex
\documentclass[11pt]{article}
\pdfoutput=1 
\usepackage{./aug/jheppub}

\usepackage{graphicx}
\usepackage{amsfonts,amsmath,amssymb}
\usepackage{tikz}
\usepackage{longtable}
\usepackage{verbatim}
\usepackage{array,setspace,mathrsfs,yfonts,dsfont,bbm,colonequals,amscd}
\usepackage{relsize,suffix,mathtools,cancel,bbm}

\usepackage{multirow}
\usepackage{longtable}

\input{./aug/aug}

\input{titlepage}

\begin{document} 

\maketitle
\flushbottom


\input{sections/introduction}

\

\input{sections/background}

\

\input{sections/fourdim}

\

\input{sections/sixdimnew}


\section*{Acknowledgments}
The authors are grateful to Shlomo S. Razamat for helpful discussions.  GZ is supported in part by World Premier International Research Center Initiative (WPI), MEXT, Japan. The research of OS is supported by the Israel Science Foundation under grant No.~2289/18, by the I-CORE Program of the Planning and Budgeting Committee, and by the Daniel scholarship for PhD students.   \vfill\eject


\appendix

\input{sections/appendix}

\bibliographystyle{./aug/ytphys}
\bibliography{./aug/refs}

\end{document}

%% file: aug/aug.tex
\newcommand{\be}{\begin{eqnarray}}
\newcommand{\ee}{\end{eqnarray}}
\newcommand{\bea}{\begin{eqnarray}}
\newcommand{\eea}{\end{eqnarray}}

\newcommand{\bn}{\begin{enumerate}}
\newcommand{\en}{\end{enumerate}}



%% file: titlepage.tex
\title{Symmetry enhancement in 4d Spin(n) gauge theories and compactification from 6d}

\preprint{}

\author[a]{Orr Sela \!}
\author[b]{and Gabi Zafrir \!}

\affiliation[a]{Department of Physics, Technion, Haifa, 32000, Israel}
\affiliation[b]{Kavli IPMU (WPI), UTIAS, the University of Tokyo, Kashiwa, Chiba 277-8583, Japan}

\emailAdd{sorrsela@campus.technion.ac.il}
\emailAdd{gabi.zafrir@ipmu.jp}

\abstract
{
We consider a known sequence of dualities involving $4d$ ${\cal N}=1$ theories with $Spin(n)$ gauge groups and use it to construct a new sequence of models exhibiting IR symmetry enhancement. Then, motivated by the observed pattern of IR symmetries we conjecture six-dimensional theories the compactification of which on a Riemann surface yields the $4d$ sequence of models along with their symmetry enhancements, and put them to several consistency checks. 
}

%% file: sections/introduction.tex
\section{Introduction}
\label{int}

Renormalization group flows of quantum field theories can entail phenomena that appear to be nontrivial from the point of view of the UV theory. One such phenomenon is symmetry enhancement, in which certain operators become conserved currents at the IR fixed point, thereby making the global symmetry in the IR larger than the one present in the UV. Even though many examples of such enhancements are known, we still lack a general explanation of them from first principles. A natural direction towards this goal would be to try to find an organizing principle for some of the known examples, and look for a systematic way to obtain new ones.

Another nontrivial phenomenon is that of duality, where two different UV models flow to the same IR theory. Similarly to the case of symmetry enhancement, there are many known examples of dualities though an understanding of the general mechanism behind it is missing. In some cases, there is a relation between these two kinds of effects, and a given duality can direct us to a model in which the symmetry enhances, and vice versa. This may happen in more than one way, and the simplest model which demonstrates all these possibilities is the four dimensional $\mathcal{N}=1$ $SU(2)$ SQCD model with four flavors, that is eight chiral fields in the fundamental representation. In this theory, since the fundamental representation of the group $SU(2)$ is pseudoreal, there is no real distinction between quarks and anti-quarks and the flavor symmetry is in fact $SU(8)$ rather than $SU(4)\times SU(4)\times U(1)$. Moreover, this model is known to have a Seiberg-dual description \cite{Seiberg:1994pq} in terms of another $SU(2)$ gauge theory with four flavors, this time with the addition of a gauge-singlet field coupled through a superpotential to the mesons of the theory. In this dual model, the presence of the gauge-singlet field keeps the UV global symmetry as $SU(4)\times SU(4)\times U(1)$, and due to the duality we conclude that it enhances in the IR to $SU(8)$. We therefore see that a duality can be used to infer that the symmetry of a model enhances in the IR, given that the dual description possesses an enlarged symmetry. The opposite direction might also be useful, and in some cases an enhancement of symmetry can suggest a new dual description in which the enhanced symmetry is manifest in the UV, see \cite{Fazzi:2018rkr} for a recent example. 

Another, less trivial instance in which certain dualities and models with symmetry enhancement are related was discussed in \cite{Razamat:2017wsk,Razamat:2018gbu} (see also \cite{Wang:2017txt}). In this case, we focus on a specific kind of dualities, called "self dualities", in which the gauge sector of the two theories is the same and they may differ only by gauge invariant fields. The $SU(2)$ Seiberg duality discussed above is an example for such a duality, in which the two dual theories have the same gauge group and matter in the fundamental representation, and differ only by the gauge-singlet fields that are coupled to the mesons in the superpotential. As discussed in \cite{Razamat:2017wsk,Razamat:2018gbu}, in many cases where two or more models are self dual it is possible to construct a closely related theory which is mapped under the duality transformation to itself ({\it i.e.} without any extra gauge-singlet fields) but with a nontrivial action on the global symmetry. Then, in some cases, the duality will act as an additional symmetry operation, resulting in an enhancement of the symmetry in the IR. In \cite{Razamat:2017wsk} it was shown how the $SU(2)$ self duality discussed above leads to a new $SU(2)$ gauge theory in which the UV symmetry is $SU(2)\times SU(6)\times U(1)$ and it enhances in the IR to $E_{6}\times U(1)$. The conformal manifold of this theory is composed of a single point, at which the duality acts as an extra symmetry\footnote{Let us comment that another variant of this kind of relations between dualities and symmetry enhancement has been discussed in \cite{Dimofte:2012pd,Razamat:2017hda}. In this case, one uses two copies of the $SU(2)$ gauge theory (and a suitable exactly marginal deformation) in order to construct a model which is mapped to itself under the duality transformation, and in which the symmetry enhances in the IR. This is different from the construction of \cite{Razamat:2017wsk} which involves only one copy, and results in an enhancement of the symmetry from $SU(8)$ in the UV to $E_7\times U(1)$ in the IR.}. 

Notice that this relation can also be applied in the other direction, and in some cases a model with symmetry enhancement can be used to find a new self-duality, see \cite{Razamat:2017wsk,Razamat:2018gbu} for explicit examples.  

In this paper we will consider a sequence of ${\cal N}=1$ self dualities with $Spin(n+8)$ ($1\leq n\leq6$) gauge groups, first appeared in \cite{Karch:1997jp}, and use it as mentioned above to construct a new sequence of models with symmetry enhancement. Then, by identifying the resulting pattern of IR symmetries (given in Eq. \eqref{4dpatt8}) we will be able to conjecture its origin from six-dimensional geometric constructions. Explicitly, we will conjecture that the $4d$ fixed points along with their enhanced symmetries can be generated through the compactification of a family of $6d$ SCFTs on a Riemann surface with fluxes for the global symmetries (see \cite{Gaiotto:2009we,Benini:2009mz,Bah:2012dg,Gaiotto:2015usa,Razamat:2016dpl,Bah:2017gph,Kim:2017toz,Kim:2018bpg,Kim:2018lfo,Razamat:2018gro,Zafrir:2018ss,Chen:2019njf,PRSZ} for discussions of such compactifications). The motivation here is to seek a physical explanation for the enhancement, in which these fixed points are viewed as low-energy effective $4d$ theories obtained from putting $6d$ SCFTs on a compact Riemann surface with fluxes. The $4d$ IR symmetries are then just the ones inherited from the corresponding $6d$ SCFTs (and preserved by the fluxes). This analysis is analogous to the one performed in \cite{Razamat:2018gbu} for the sequence of self dualities with $Spin(n+4)$ ($1\leq n\leq8$) gauge groups (first appeared in \cite{Csaki:1997cu}), and extends it to the $Spin(n+8)$ sequence mentioned above. In fact, we find that the IR symmetry patterns and the $6d$ theories corresponding to the two sequences are related to each other in a simple way. 

As we are focusing on four dimensional theories with ${\cal N}=1$ supersymmetry, we will be able to employ certain non-perturbative techniques in order to extract exact information about these theories at their strongly-coupled IR fixed points. Our main tool will be the superconformal index \cite{Romelsberger:2005eg,Kinney:2005ej,Dolan:2008qi,Rastelli:2016tbz}, which due to its invariance under the renormalization group flow \cite{Festuccia:2011ws,Dumitrescu:2016ltq} enables us to easily extract the IR conserved currents of a theory by a simple calculation in the UV, where the theory is weakly coupled. Then, by performing such a calculation for each one of the theories in the sequence, we can find their IR symmetries and look for a possible pattern. As mentioned above, this can then suggest an alternative description of these IR models in which they are viewed as effective $4d$ theories obtained from putting six-dimensional theories on a compact Riemann surface with fluxes.  

The organization of this paper is as follows. We begin in section \ref{back} with a review of the superconformal index and its computation, and briefly describe the results of \cite{Razamat:2018gbu} that will be useful for the rest of the paper. Next, in section \ref{fourdim} we construct the sequence of $Spin(n+8)$ models with symmetry enhancement and find the IR symmetry of each model. After identifying the resulting pattern, we turn in section \ref{sixdim} to discuss the six-dimensional theories the compactification of which on a Riemann surface is conjectured to yield the $4d$ sequence along with its symmetry enhancements. An accompanying appendix includes an example for a detailed computation of the superconformal index for an interesting model with symmetry enhancement that was not discussed in section \ref{fourdim}.

%% file: sections/background.tex
\section{Background}
\label{back}

In this section we review some of the ingredients and previous results needed for the analysis in this paper. We begin in subsection \ref{index} with a review of the superconformal index and its properties that will be used repeatedly in later sections. Then, we continue in subsection \ref{np4} with recalling some of the results presented recently in \cite{Razamat:2018gbu} which will be relevant for our discussion. 

\subsection{The superconformal index}
\label{index}

The index of a four dimensional superconformal field theory can be defined as the Witten index of the theory in radial quantization \cite{Romelsberger:2005eg,Kinney:2005ej,Dolan:2008qi,Rastelli:2016tbz}. That is, it can be written as a trace over the states of the theory quantized on ${\mathbb S}^3\times {\mathbb R}$. Alternatively, the index can be defined as the supersymmetric partition function on ${\mathbb S}^3\times {\mathbb S}^1$, where localization techniques can be employed for its computation \cite{Closset:2013sxa,Assel:2014paa}. In this paper, we focus on the former definition; then, denoting by $\cal Q$ one of the Poincar\'e supercharges\footnote{More explicitly, we choose here $\cal Q\equiv \widetilde{\mathcal{Q}}_{\dot{-}}$ which in the language of \cite{Rastelli:2016tbz} corresponds to the "right-handed index".}, the index is written as the following weighted trace: 
\begin{equation}
\label{indtrace}
{\cal I}\left(p,q;{u_a}\right)=\mathrm{Tr}_{{\mathbb S}^3}\left[(-1)^F e^{-\beta \delta} p^{j_1+j_2+\frac{r}{2}} q^{j_2-j_1+\frac{r}{2}} \prod_{a}u_{a}^{e_{a}} \right]\,.
\end{equation}
Here $F$ is the fermion number, $p$ and $q$ are the superconformal fugacities, $j_1$ and $j_2$ are the Cartan generators of the $SU(2)_1\times SU(2)_2$ isometry group of ${\mathbb S}^3$ (the indices of the corresponding doublets are denoted by $\pm$ and $\dot{\pm}$, respectively), $r$ is the $U(1)_r$ charge, and $u_{a}$ and $e_{a}$ are the fugacities and charges of the global symmetries, respectively. Moreover, $\delta$ is defined as the following anti-commutator: 
\begin{equation}
\label{delta}
\delta\equiv\left\{ \mathcal{Q},\mathcal{Q}^{\dagger}\right\} =E-2j_{2}-\frac{3}{2}r\,.
\end{equation}
Note that even though the chemical potential $\beta$ appears in the definition \eqref{indtrace}, the index is in fact independent of it since the states with $\delta>0$ come in boson/fermion pairs and therefore cancel. Only the states with $\delta=0$, corresponding to short multiplets of the superconformal algebra, contribute.

\

The most salient feature of the index, which also applies to some other partition functions, is its invariance under the RG flow \cite{Festuccia:2011ws,Dumitrescu:2016ltq}. Hence, by calculating the index of an asymptotically free theory at the UV fixed point, we automatically obtain the index of this theory in the IR where the interactions might be strong. For that purpose, this calculation, even though performed in the UV, involves the IR R charge, {\it i.e.} the one that appears in the superconformal algebra of the theory at the IR fixed point. The only assumptions in such a calculation are that we flow to an interacting fixed point without free fields, and that we have identified the IR R symmetry correctly.  We next outline the calculation of the index at the UV fixed point, referring to \cite{Rastelli:2016tbz} for more details.  

Our starting point is the trace over the Hilbert space of the theory. Since we consider a conformal theory, we can employ the state/operator correspondence and instead of summing over the states of the theory we can list all the gauge invariant operators one can build from the different fields and sum their contributions to the index according to \eqref{indtrace}. As discussed above, only the contributions of the operators with $\delta=0$ (where $\delta$ is given by \eqref{delta}) do not cancel and therefore we can restrict to counting them only. 

The advantages of performing the calculation in the UV (where the theory is free) are that the dimensions of the various fields are given by their classical values ({\it i.e.} they do not have anomalous dimensions) and that the dimensions of composite operators equal to the sum of the dimensions corresponding to the constituent fields. As a result, an operator will have $\delta=0$ only if each of the fields used to build it has $\delta=0$. Therefore, we can restate the counting procedure as follows: Using only the modes of the fields with $\delta=0$ \footnote{Note that in addition to the fields themselves, one also uses the spacetime derivatives with $\delta=0$, as discussed below.}, we build all the possible gauge invariant operators and sum their contributions to the index. 

We are now only left with finding the fields with $\delta=0$. For a chiral multiplet $\Phi$ with an IR R charge $r$, those are the scalar $\Phi$ (denoted in the same way as the multiplet) and the anti-spinor component $\bar{\psi}_{\dot{+}}^{\Phi}$. The contribution of the scalar to the index according to \eqref{indtrace} is $+\left(pq\right)^{\frac{r}{2}}$ while that of the anti-spinor is $-\left(pq\right)^{\frac{2-r}{2}}$. Note that in this discussion, the parts corresponding to the gauge and global symmetries are suppressed. Next, among the fields that belong to a vector multiplet, only the gaugino $\lambda$ has $\delta=0$. The component $\lambda_{+}$ contributes $-p$ to the index while $\lambda_{-}$ contributes $-q$. In addition to the aforementioned fields, also the two derivatives $\partial_{\pm\dot{+}}$ have $\delta=0$, and $\partial_{+\dot{+}}$ contributes $+p$ to the index while $\partial_{-\dot{+}}$ contributes $+q$. Finally, the gaugino equation of motion $\partial_{-\dot{+}}\lambda_{+}+\partial_{+\dot{+}}\lambda_{-}=0$ has itself $\delta=0$ and should be taken into account, with a contribution of $+2pq$ \footnote{Notice that there is a plus sign here instead of a minus because the contribution of this equation should be removed from the index and not added. Note also that the equations of motion of the other fields mentioned before have $\ensuremath{\delta\neq0}$ and so do not contribute to the index.}. Now that we have all the ingredients, we can construct all the possible gauge invariant operators with $\delta=0$ and calculate the index. A detailed example for such a calculation can be found in the appendix. 

\

At this stage, once we have the index for a given model, a natural question to raise is how to extract physical information about the IR theory from it. In \cite{Beem:2012yn}, it was shown that the expansion of the index in powers of $p$ and $q$ holds in a simple way information about the relevant and marginal operator content of the theory, as well as on the symmetry in the IR. Explicitly, it was shown that the chiral primary operators that contribute to the index at order $(pq)^{r/2}(p/q)^0$ ($r<2$) are in fact the only kind of operators that contribute at this order. As a result, these operators, which are precisely the relevant deformations of the IR SCFT, can be easily identified from the expansion of the index by analyzing the coefficients at orders $(pq)^{r/2}(p/q)^0$ ($r<2$). In addition, it was shown that the contribution at order $pq(p/q)^0$ consists of the marginal operators that contribute with a positive sign and the conserved currents that contribute with a negative sign. This form allows for possible cancellations between these two kinds of operators, and reflects the phenomenon discussed in detail in \cite{Green:2010da}, according to which marginal operators can fail to be exactly marginal if and only if they combine with currents that correspond to broken global symmetries to form long multiplets. The index in general only counts short multiplets up to such recombinations into long ones, hence such long multiplets can not contribute to the index and so marginal and conserved current operators can only appear together and with an opposite sign. Notice also that since the total contribution of such recombinations is zero, the index is constant on the conformal manifold (if there is one). Summarizing the discussion in this paragraph, the expansion of the index in powers of $p$ and $q$ assumes the following form,   
\begin{equation}
\label{indform}
\mathcal{I}=1+\left(\textrm{Relevants}\right)\left(pq\right)^{\#<1}+\left(\textrm{Marginals-Currents}\right)pq+\ldots
\end{equation}

As mentioned, finding the coefficient at order $pq$ is usually very useful in determining the symmetry of a theory in the IR, and we will use it repeatedly in this paper. In some preferable cases, this coefficient turns out to be negative, and we can conclude that the symmetry in the IR is at least the one for which (minus) the adjoint representation reproduces this coefficient. Note also that in general, if one is interested only in the index at order $pq$ then the result can be obtained by listing the operators that contribute at this order and summing their contributions, without performing the calculation of the entire index. A detailed example for such a calculation can be found in the appendix. 

\

\subsection{Symmetry enhancement in Spin(n+4) models and compactification from 6d}
\label{np4}

In this subsection we briefly review some of the results presented recently in \cite{Razamat:2018gbu}, which are closely related to the analysis of this paper and share the same logic. We examine a sequence of theories, each one of which is known to have at least one self-dual description, that is a dual model with the same gauge sector and possibly with extra gauge-singlet fields and a superpotential. Then, motivated by these self dualities, we deform each theory by a relevant superpotential such that under the corresponding duality the theory is mapped to itself but with a nontrivial action on the global symmetry it possesses. The theories we obtain in this way have a larger symmetry in the IR than the one present in the UV, and we get a pattern of emergent symmetries. One natural direction to look for the origin of this pattern is to examine compactifications of six dimensional theories on a Riemann surface with fluxes for the global symmetries, as recently explained in \cite{Kim:2017toz,Kim:2018bpg,Kim:2018lfo,Zafrir:2018ss}. And indeed, using such compactifications one can argue for the four dimensional enhanced symmetries we encounter, as discussed below. Let us next present this procedure in more detail.  

We consider a sequence of theories, first appeared in \cite{Csaki:1997cu}, with $Spin(n+4)$ ($1\leq n\leq8$) gauge groups, each one containing $n$ chiral fields in the vector representation along with chiral fields in the spinor representations (both chiralities) with a total of 32 components. Each such theory has at least one self-dual description, and by deforming it as discussed above we construct a model with a nontrivial symmetry enhancement. In most of these models, the dimension of the conformal manifold vanishes and the IR currents are identified easily from the coefficient in the expansion of the index at order $pq$ [recall Eq. \eqref{indform}]. In the other cases, there is a nontrivial conformal manifold but the symmetry can still be found by examining the index at order $pq$ as well as other orders and checking that all the representations form characters of the ones of the larger symmetry group. 

We list the models with symmetry enhancement constructed from the $Spin(n+4)$ self-duality sequence of \cite{Csaki:1997cu} in the table below. Here the gauge-charged matter content is denoted by (Number of spinors, Number of vectors) for odd $n$ and by (Number of spinors of one chirality, Number of spinors of the other chirality, Number of vectors) for even $n$, while the gauge-singlet fields are denoted by $F_i$\footnote{Throughout this note, gauge and global symmetry indices will be suppressed in the superpotentials.}. We denote the chiral fields in the vector representation by $V$, and those in the spinor representations by $S$ and $C$ for the two chiralities. 
\begin{center}
	\begin{tabular}{|c|c|c|}
		\hline
		Gauge group & Gauge-charged matter content & Superpotential\\
		\hline\hline
		$Spin\left(5\right)$ & $\left(8,1\right)$ & $F_0 S^{2}+F_1 V^{2}$\\
		\hline
		$Spin\left(6\right)$ & $\left(4,4,2\right)$ & $F_0 CV^{2}S+F_1 VC^{2}+F_2 V^{2}$\\
		\hline
		$Spin\left(7\right)$ & $\left(4,3\right)$ & $F_0S^{2}V+F_1S^{2}+F_2V^{2}$\\
		\hline 
		$Spin\left(8\right)$ & $\left(2,2,4\right)$ & $F_0SCV+F_1S^{2}+F_2C^{2}+F_3V^{2}$\\
		\hline 
		$Spin\left(9\right)$ & $\left(2,5\right)$ & $F_0S^{2}V^{2}+F_1S^{2}V+F_2S^{2}+F_3V^{2}$\\
		\hline 
		$Spin\left(10\right)$ & $\left(2,0,6\right)$ & $F_0S^{2}V+F_1V^{2}$\\
		\hline 
		$Spin\left(11\right)$ & $\left(1,7\right)$ & $F_0S^{2}V^{2}+F_1S^{2}V+F_2V^{2}$\\
		\hline 
		$Spin\left(12\right)$ & $\left(1,0,8\right)$ & $F_0S^{2}V^{2}+F_1V^{2}$\\
		\hline
	\end{tabular}
\end{center}

The symmetry enhancement observed in each of the models above is as follows \cite{Razamat:2018gbu}, 
\begin{center}
	\begin{tabular}{|c|c|c|c|c|}
		\hline
		Gauge group & UV  symmetry & IR  symmetry & UV rank & IR rank\\
		\hline	\hline  
		$Spin\left(5\right)$ & $SU\left(8\right)\times U\left(1\right)$ & $E_{7}\times U\left(1\right)$ & 8 & 8\\
		\hline 
		$Spin\left(6\right)$ & $SU\left(2\right)\times SU\left(4\right)^{2}\times U\left(1\right)^{2}$  & $SU\left(2\right)\times SO\left(12\right)\times U\left(1\right)^{2}$ & 9 & 9\\
		\hline 
		$Spin\left(7\right)$ & $SU\left(4\right)\times SU\left(3\right)\times U\left(1\right)$ & $SU\left(6\right)\times SU\left(3\right)^{2}\times U\left(1\right)$ & 6 & 10\\
		\hline 
		$Spin\left(8\right)$ & $SU\left(2\right)^{2}\times SU\left(4\right)\times U\left(1\right)^{2}$ & $SU\left(4\right)^{3}\times U\left(1\right)\times SU(2)$ & 7 & 11\\
		\hline 
		$Spin\left(9\right)$ & $SU\left(2\right)\times SU\left(5\right)\times U\left(1\right)$ & $SU\left(3\right)\times SU\left(5\right)^{2}\times U\left(1\right)^{2}$ & 6 & 12\\
		\hline 
		$Spin\left(10\right)$ & $SU\left(2\right)\times SU\left(6\right)\times U\left(1\right)$ & $SU\left(3\right)\times SU\left(6\right)^{2}\times U\left(1\right)$ & 7 & 13\\
		\hline 
		$Spin\left(11\right)$ & $SU\left(7\right)\times U\left(1\right)$ & $SU\left(7\right)^{2}\times U\left(1\right)^{2}$ & 7 & 14\\
		\hline 
		$Spin\left(12\right)$ & $SU\left(8\right)\times U\left(1\right)$ & $SU\left(8\right)^{2}\times U\left(1\right)$ & 8 & 15\\
		\hline
	\end{tabular}
	\label{table:Summary}
\end{center}

We identify the following pattern of IR symmetries:
\begin{equation}
\label{4dpatt}
\textrm{4d IR symmetry:}\,\,\,\,\,\,\,C_{E_{9-n}}\left(SU\left(2\right)\right)\times SU\left(n\right)^{2}\times U\left(1\right)
\end{equation}
where $C_{H}(G)$ is the commutant of $H$ in $G$ and $E_{9-n}$ commonly denotes the commutant of $SU(n)$ in $E_8$, {\it i.e.} $E_{9-n}=C_{SU(n)}\left(E_8\right)$. As a result, we can also write: 
\begin{equation}
\textrm{4d IR symmetry:}\,\,\,\,\,\,\,C_{E_{8}}\left(SU\left(n\right)\times SU\left(2\right)\right)\times SU\left(n\right)^{2}\times U\left(1\right).
\end{equation}
Note that in the $Spin(6)$ case ($n=2$), we see the part $SU\left(2\right)\times U\left(1\right)^{2}$ instead of $SU\left(2\right)^{2}\times U\left(1\right)$, that is one of the two $SU(2)$ groups is further broken to a $U(1)$. We will comment on this below.   

\

Once the pattern of IR symmetries has been identified, we may turn to investigate its origin. One way to do it is to try to find a theory in six dimension and a certain compactification that produce the $4d$ models along with their enhanced symmetries in the IR. These symmetries, in turn, will be the subgroups of the $6d$ ones preserved by the compactification. 

An example for this kind of reasoning was given in \cite{Kim:2017toz}, where it was shown how the $E_7$ surprise theory of \cite{Dimofte:2012pd} with the $E_7$ enhancement can be obtained by compactifying (and mass-deforming) the rank 1 E-string theory, a $6d$ (1,0) model with $E_8$ symmetry, on a torus with fluxes that break the $E_8$ to $E_7\times U(1)$. This theory can be considered as the $Spin(4)$ ($n=0$) case in the $Spin(n+4)$ sequence we consider, and so it is reasonable to expect a similar construction for the other symmetry enhancements appearing in the pattern. 

The first stage in finding such a construction would be to find the $6d$ theories the compactification of which will lead to the $Spin(n+4)$ sequence. As suggested in \cite{Kim:2017toz,Kim:2018bpg,Kim:2018lfo}, we can do it by considering the $5d$ versions of the $4d$ theories\footnote{By the $5d$ versions we mean taking the matter content of the $4d$ ${\cal N} = 1$ theories and replacing it with the $5d$ ${\cal N}=1$ one, so that $4d$ chiral fields become $5d$ hypermultiplets and so forth.}, investigate their UV behavior and check whether there is a $6d$ completion of them, that is $6d$ SCFTs that when compactified on a circle yield these $5d$ theories. Doing this for the $4d$ $Spin(n+4)$ theories while using \cite{Zafrir:2015uaa} for determining the global symmetries at the $5d$ UV fixed points, we indeed find that the $5d$ theories have a UV completion in terms of $6d$ theories that have the global symmetry $SU(2n)\times E_{9-n}$, where $E_{9-n}$ was defined below \eqref{4dpatt}.

This leads us to conjecture that the $6d$ SCFTs we are looking for can be engineered in the following way. We take the rank 1 E-string theory and gauge an $SU(n)$ subgroup of $E_8$, while also adding $2n$ hypermultiplets in the fundamental representation of $SU(n)$. With this combination, the $SU(n)$ gauge anomaly can be canceled by adding a tensor multiplet as is usual in $6d$ low-energy gauge theory descriptions of $6d$ SCFTs. Then, the infinite coupling limit would correspond to the origin of the tensor branch of some $6d$ SCFT, which is the one that we are looking for.

Now that we have found the $6d$ theories, we conjecture that the combined effect of compactifying them on a torus with fluxes and adding an appropriate relevant deformation yields the expected $4d$ theories in which the $SU(2n)$ part of the $6d$ global symmetry is broken to $SU(n)^2\times U(1)$ (and in the $n=2$ case one of the $SU(2)$s is further broken to $U(1)$) and the $E_{9-n}$ part is broken to the commutant of $SU(2)$ in it. This procedure is demonstrated explicitly in \cite{Razamat:2018gbu} for the $n=1$ and $n=2$ cases of the $Spin(n+4)$ sequence and provides the desired $6d$ construction for the observed symmetry enhancement pattern.

%% file: sections/fourdim.tex
\section{Four dimensional symmetry enhancement in Spin(n+8) models}
\label{fourdim}

We consider a sequence of theories which is reminiscent of the $Spin(n+4)$ sequence discussed in \cite{Razamat:2018gbu} and reviewed above in subsection \ref{np4}. Specifically, we examine a sequence of theories, first appeared in \cite{Karch:1997jp}, with $Spin(n+8)$ ($1\leq n\leq6$) gauge groups\footnote{The case $n=0$, corresponding to a $Spin(8)$ gauge group, will be discussed in the appendix.}, each one containing $n$ chiral fields in the vector representation along with chiral fields in the spinor representations (both chiralities) with a total of 64 components. As in the $Spin(n+4)$ sequence, each theory has at least one self-dual description \cite{Karch:1997jp} and by adding free gauge-singlet fields and deforming the theory by an appropriate relevant superpotential, we can obtain a model in which the symmetry enhances in the IR. Then, collecting the IR symmetries we observe in the various models, a specific pattern can be identified which can be later matched with a six dimensional construction, as demonstrated before. 

We next turn to examine each one of the theories with symmetry enhancement in the $Spin(n+8)$ sequence, presenting the matter contents and superpotentials used for their definition along with the parts of the corresponding superconformal indices needed for extracting their IR symmetries. In the $n=6$ case, we analyze two models that differ by the choice of superpotential; one choice yields an emergent IR symmetry that fits the pattern we observe in the sequence, while the other corresponds to an even larger enhancement. We close the section with a summary of the various IR symmetries and use it to identify a pattern.   

Let us comment that the superpotentials in the models we discuss in this paper are built from gauge-singlet fields $F_i$ that are coupled linearly to certain composite operators, thereby setting them to zero in the chiral ring. This form of coupling is usually referred to in the literature as "flipping" of the corresponding composite operators, and we use it in our analysis for three different reasons. The first is to obtain a model with symmetry enhancement, as mentioned before and as discussed in detail in \cite{Razamat:2018gbu} (see especially section 5 on ring relations and hidden symmetries). The second is to prevent having operators that violate the unitarity bound, as elaborated in \cite{Benvenuti:2017lle}. Such violating operators will then become free in the IR, meaning that there are accidental $U(1)$ symmetries and that we have not identified the superconformal R symmetry correctly. The third reason that we might use such a flipping is to remove extra marginal operators from the spectrum of the IR theory. This enables us to read the IR symmetry from the index in an easier way, and in some cases to obtain a model in which we can prove that the symmetry indeed enhances [see the discussion below Eq. \eqref{indform} and in subsection \ref{np4}].  

\

\subsection*{Spin(9)}

\

We begin with the first theory in the sequence, corresponding to $n=1$. The matter content is given in the following table. 
\begin{center}
	\begin{tabular}{|c||c|c|c|c|}
		\hline
		Field & $Spin\left(9\right)_{g}$ & $SU\left(4\right)_{s}$ & $U\left(1\right)_{a}$ & $U\left(1\right)_{r}$\\
		\hline 
		
		$S$ & $\boldsymbol{16}$ & $\boldsymbol{4}$ & 1 & $\frac{1}{4}$\\
		
		$V$ & $\boldsymbol{9}$ & $\boldsymbol{1}$ & -8 & 0\\
		
		$F_{0}$ & $\boldsymbol{1}$ & $\boldsymbol{\overline{10}}$ & 6 & $\frac{3}{2}$\\
		
		$F_{1}$ & $\boldsymbol{1}$ & $\boldsymbol{20'}$ & -4 & 1\\
		
		$F_{2}$ & $\boldsymbol{1}$ & $\boldsymbol{\overline{10}}$ & -2 & $\frac{3}{2}$\\
		
		$F_{3}$ & $\boldsymbol{1}$ & $\boldsymbol{1}$ & 16 & 2\\
		\hline
	\end{tabular}
	\label{table:Spin9}
\end{center}
The superpotential is given by 
\begin{equation}
W=F_{0}S^{2}V+F_{1}S^{4}+F_{2}S^{2}+F_{3}V^{2}
\end{equation}
and to find the superconformal R charge we use $a$ maximization \cite{Intriligator:2003jj} and get $\hat{r}=r-0.0188q_{a}$, where $q_{a}$ is the charge under $U\left(1\right)_{a}$. Note also that all the gauge-invariant operators are above the unitarity bound and the superpotential is a relevant deformation of the theory defined without it. 

Turning to find the IR symmetry of this model, we compute the index as described in subsection \ref{index} and demonstrated in the appendix, and find the following contribution at order $pq$ in the expansion of the index: 
\begin{equation}
\label{pq9}
-\boldsymbol{20'}_{SU\left(4\right)_{s}}-\boldsymbol{15}_{SU\left(4\right)_{s}}-1=-\boldsymbol{35}_{SU\left(6\right)}-1\,.
\end{equation}
Here and in what follows, we denote by $\boldsymbol{R}_{G}$ the character of the representation $\boldsymbol{R}$ of the group $G$. We notice that we get a negative coefficient, and that in addition to the contribution of the UV currents $\boldsymbol{15}_{SU\left(4\right)_{s}}$ and 1 (corresponding to the $SU\left(4\right)_{s}\times U\left(1\right)_a$ UV global symmetry) we have the additional piece $\boldsymbol{20'}_{SU\left(4\right)_{s}}$ that combines with the adjoint of $SU\left(4\right)_{s}$ to form the adjoint representation of the larger group $SU(6)$. Following our previous discussion [see Eq. \eqref{indform}] we conclude that the IR symmetry is at least $SU\left(6\right)\times U\left(1\right)$. As we do not have any evidence for an even larger symmetry or for accidental $U(1)$s, we conclude that the UV symmetry $SU\left(4\right)\times U\left(1\right)$ enhances to $SU\left(6\right)\times U\left(1\right)$ in the IR. Moreover, as we do not have any contribution from marginal operators, we infer that the dimension of the conformal manifold vanishes, {\it i.e.} that it is composed of a single point. 

As a consistency check for this enhancement of symmetry, we can consider other parts of the index and check whether the representations of $SU\left(4\right)_{s}$ form representations of $SU\left(6\right)$. We indeed find that we can write the various powers in terms of characters of $SU\left(6\right)$, as demonstrated in the following expansion which also includes the contributions coming from relevant operators: 
\begin{equation*}
\mathcal{I}=1+\left(\boldsymbol{10}_{SU\left(4\right)_{s}}+\boldsymbol{\overline{10}}_{SU\left(4\right)_{s}}\right)\left(a^{6}+a^{-2}\right)\left(pq\right)^{\frac{3}{4}}+\left(\boldsymbol{15}_{SU\left(4\right)_{s}}+\boldsymbol{20'}_{SU\left(4\right)_{s}}\right)a^{-4}\left(pq\right)^{\frac{1}{2}}+
\end{equation*}
\begin{equation*}
+\left(a^{16}-\boldsymbol{15}_{SU\left(4\right)_{s}}-\boldsymbol{20'}_{SU\left(4\right)_{s}}-1\right)pq+\ldots
\end{equation*}
\begin{equation}
\label{ind9}
=1+\boldsymbol{20}_{SU\left(6\right)}\left(a^{6}+a^{-2}\right)\left(pq\right)^{\frac{3}{4}}+\boldsymbol{35}_{SU\left(6\right)}a^{-4}\left(pq\right)^{\frac{1}{2}}+\left(a^{16}-\boldsymbol{35}_{SU\left(6\right)}-1\right)pq+\ldots
\end{equation}
where $a$ is the fugacity corresponding to $U(1)_a$.

It is important to comment that since $U(1)_a$ mixes with $U(1)_r$ in the expression for the superconformal R charge, the contribution $a^{16}pq$ that appears in \eqref{ind9} does not correspond to a marginal operator; indeed, it corresponds to the relevant operator $F_3$. In other words, the index \eqref{ind9} is written using $U(1)_r$ instead of $U(1)_{\hat{r}}$ (as it is clearer due to the irrational mixing) and one should remember the mixing with $U(1)_a$. The marginal operators and conserved currents only contribute at order $pq$ when the true superconformal R-symmetry is used. In the cases studied here, for computational ease, we shall generally use an R-symmetry that is not the superconformal R-symmetry but under which the R-charge of the fields are rational. The true superconformal R-symmetry is then a mixture of it with flavor symmetries with irrational coefficients. In such cases when we want to find the IR symmetry using the coefficient at order $pq$, we should restrict only to contributions with a vanishing $U(1)_a$ charge, as in \eqref{pq9}. Only such contributions correspond to superconformal R charge which is equal to 2, and therefore to marginal and conserved current operators. 

\

\subsection*{Spin(10)}

\

We next turn to the $n=2$ case, with matter content as follows, 
\begin{center}
	\begin{tabular}{|c||c|c|c|c|c|c|c|}
		\hline
		Field & $Spin\left(10\right)_{g}$ & $SU\left(2\right)_{s}$ & $SU\left(2\right)_{c}$ & $SU\left(2\right)_{v}$ & $U\left(1\right)_{a}$ & $U\left(1\right)_{b}$ & $U\left(1\right)_{r}$\\
		\hline 
		
		$S$ & $\boldsymbol{16}$ & $\boldsymbol{2}$ & $\boldsymbol{1}$ & $\boldsymbol{1}$ & 1 & 0 & $\frac{1}{4}$\\
		
		$C$ & $\boldsymbol{\overline{16}}$ & $\boldsymbol{1}$ & $\boldsymbol{2}$ & $\boldsymbol{1}$ & 0 & 1 & $\frac{1}{4}$\\
		
		$V$ & $\boldsymbol{10}$ & $\boldsymbol{1}$ & $\boldsymbol{1}$ & $\boldsymbol{2}$ & -2 & -2 & 0\\
		
		$F_{0}$ & $\boldsymbol{1}$ & $\boldsymbol{2}$ & $\boldsymbol{2}$ & $\boldsymbol{2}$ & -1 & 1 & 1\\
		
		$F_{1}$ & $\boldsymbol{1}$ & $\boldsymbol{2}$ & $\boldsymbol{2}$ & $\boldsymbol{1}$ & -1 & -1 & $\frac{3}{2}$\\
		
		$F_{2}$ & $\boldsymbol{1}$ & $\boldsymbol{1}$ & $\boldsymbol{3}$ & $\boldsymbol{2}$ & 2 & 0 & $\frac{3}{2}$\\
		
		$F_{3}$ & $\boldsymbol{1}$ & $\boldsymbol{1}$ & $\boldsymbol{1}$ & $\boldsymbol{3}$ & 4 & 4 & 2\\
		
		$F_{4}$ & $\boldsymbol{1}$ & $\boldsymbol{1}$ & $\boldsymbol{1}$ & $\boldsymbol{1}$ & 0 & -4 & 1\\
		
		$F_{5}$ & $\boldsymbol{1}$ & $\boldsymbol{1}$ & $\boldsymbol{1}$ & $\boldsymbol{1}$ & -4 & 0 & 1\\
		
		$F_{6}$ & $\boldsymbol{1}$ & $\boldsymbol{3}$ & $\boldsymbol{1}$ & $\boldsymbol{2}$ & 0 & 2 & $\frac{3}{2}$\\
		
		$F_{7}$ & $\boldsymbol{1}$ & $\boldsymbol{2}$ & $\boldsymbol{2}$ & $\boldsymbol{1}$ & 3 & 3 & $\frac{3}{2}$\\
		
		$F_{8}$ & $\boldsymbol{1}$ & $\boldsymbol{3}$ & $\boldsymbol{3}$ & $\boldsymbol{1}$ & -2 & -2 & 1\\
		\hline
	\end{tabular}
	\label{table:Spin10}
\end{center}
The superpotential is given by 
\begin{equation*}
W=F_{0}S^{3}CV+F_{1}SC+F_{2}C^{2}V+F_{3}V^{2}+F_{4}C^{4}+F_{5}S^{4}+F_{6}S^{2}V+F_{7}SCV^{2}+F_{8}S^{2}C^{2}
\end{equation*}
and the superconformal R charge by $\hat{r}=r-0.026q_{a}-0.092q_{b}$. All the gauge-invariant operators are above the unitarity bound and the superpotential is a relevant deformation of the theory defined without it. 

To find the IR symmetry of this model as we did above, we compute the index and find the following coefficient at order $pq$ (with vanishing $U(1)_a$ and $U(1)_b$ charges): 
\begin{equation*}
\textrm{Marginals}-\left(\boldsymbol{3}_{SU\left(2\right)_{s}}+\boldsymbol{3}_{SU\left(2\right)_{c}}+\boldsymbol{3}_{SU\left(2\right)_{v}}+2\right)-\left(\boldsymbol{3}_{SU\left(2\right)_{s}}\boldsymbol{3}_{SU\left(2\right)_{c}}+1\right)
\end{equation*}
\begin{equation}
\label{pq10}
=\textrm{Marginals}-\boldsymbol{15}_{SU\left(4\right)}-\boldsymbol{3}_{SU\left(2\right)_{v}}-3\,,
\end{equation}
where "Marginals" denotes positive contributions corresponding to marginal operators\footnote{These marginal operators are of the form $S^{4}C^{4}V^{2}$ and are not composed of smaller gauge-invariant operators.}, which are not written explicitly due to their length. The first parenthesis in the first line of \eqref{pq10} correspond to the UV currents, while the second ones correspond to the contribution of the extra IR currents. Note that "Marginals" does not contain the representations which appear in the second parenthesis, hence cancellations do not take place. We therefore obtain that the UV symmetry $SU\left(2\right)^{3}\times U\left(1\right)^{2}$ enhances to $SU\left(4\right)\times SU\left(2\right)\times U\left(1\right)^{3}$ in the IR. 

We can further check this enhancement in other parts of the index, as we did above in \eqref{ind9}. For example, denoting by $a$ and $b$ the fugacities of $U(1)_a$ and $U(1)_b$, we have at orders $a^2(pq)^{3/4}$ and $b^2(pq)^{3/4}$ in the expansion of the index the following coefficient:  
\begin{equation}
\left(\boldsymbol{3}_{SU\left(2\right)_{s}}+\boldsymbol{3}_{SU\left(2\right)_{c}}\right)\boldsymbol{2}_{SU\left(2\right)_{v}}=\boldsymbol{6}_{SU\left(4\right)}\boldsymbol{2}_{SU\left(2\right)_{v}}\,,
\end{equation}
while at order $a^{-1}b(pq)^{1/2}$ we have: 
\begin{equation}
2\,\boldsymbol{2}_{SU\left(2\right)_{s}}\boldsymbol{2}_{SU\left(2\right)_{c}}\boldsymbol{2}_{SU\left(2\right)_{v}}=2\,\boldsymbol{4}_{SU\left(4\right)}\boldsymbol{2}_{SU\left(2\right)_{v}}\,.
\end{equation}
Moreover, at order $a^{-2}b^{-2}(pq)^{1/2}$ we get the following contribution: 
\begin{equation}
\boldsymbol{3}_{SU\left(2\right)_{s}}\boldsymbol{3}_{SU\left(2\right)_{c}}+\boldsymbol{3}_{SU\left(2\right)_{s}}+\boldsymbol{3}_{SU\left(2\right)_{c}}+\boldsymbol{3}_{SU\left(2\right)_{v}}=\boldsymbol{15}_{SU\left(4\right)}+\boldsymbol{3}_{SU\left(2\right)_{v}}\,,
\end{equation}
and we can similarly write the other powers using representations of $SU\left(4\right)\supset SU\left(2\right)_{s}\times SU\left(2\right)_{c}$. Note that the additional $U(1)$ symmetry we have in the IR does not appear in the UV and so is not manifest in the index at powers other than $pq$. One issue with the appearance of this additional $U(1)$ is that it could in principle mix with the R-symmetry and invalidates the analysis performed here. The results of this analysis then are under the assumption that this additional $U(1)$ does not mix with the R-symmetry.   

\

\subsection*{Spin(11)}

\

The next model in the sequence ($n=3$) is as follows, 
\begin{center}
	\begin{tabular}{|c||c|c|c|c|c|}
		\hline
		Field & $Spin\left(11\right)_{g}$ & $SU\left(2\right)_{s}$ & $SU\left(3\right)_{v}$ & $U\left(1\right)_{a}$ & $U\left(1\right)_{r}$\\
		\hline 
		
		$S$ & $\boldsymbol{32}$ & $\boldsymbol{2}$ & $\boldsymbol{1}$ & -3 & $\frac{1}{4}$\\
		
		$V$ & $\boldsymbol{11}$ & $\boldsymbol{1}$ & $\boldsymbol{3}$ & 8 & 0\\
		
		$F_{0}$ & $\boldsymbol{1}$ & $\boldsymbol{3}$ & $\boldsymbol{\overline{3}}$ & -2 & $\frac{3}{2}$\\
		
		$F_{1}$ & $\boldsymbol{1}$ & $\boldsymbol{3}$ & $\boldsymbol{3}$ & -10 & $\frac{3}{2}$\\
		
		$F_{2}$ & $\boldsymbol{1}$ & $\boldsymbol{1}$ & $\boldsymbol{\overline{6}}$ & -16 & 2\\
		
		$F_{3}$ & $\boldsymbol{1}$ & $\boldsymbol{1}$ & $\boldsymbol{1}$ & 6 & $\frac{3}{2}$\\
		
		$F_{4}$ & $\boldsymbol{1}$ & $\boldsymbol{1}$ & $\boldsymbol{\overline{3}}$ & 4 & 1\\
		
		$F_{5}$ & $\boldsymbol{1}$ & $\boldsymbol{3}$ & $\boldsymbol{\overline{3}}$ & 4 & 1\\
		
		$F_{6}$ & $\boldsymbol{1}$ & $\boldsymbol{5}$ & $\boldsymbol{1}$ & 12 & 1\\
		
		$F_{7}$ & $\boldsymbol{1}$ & $\boldsymbol{1}$ & $\boldsymbol{1}$ & 12 & 1\\
		
		$F_{8}$ & $\boldsymbol{1}$ & $\boldsymbol{1}$ & $\boldsymbol{1}$ & -18 & $\frac{3}{2}$\\
		\hline
	\end{tabular}
	\label{table:Spin11}
\end{center}
The superpotential is given by 
\begin{equation*}
W=F_{0}S^{2}V+F_{1}S^{2}V^{2}+F_{2}V^{2}+F_{3}S^{2}+F_{4}S^{4}V+F_{5}S^{4}V+F_{6}S^{4}+F_{7}S^{4}+F_{8}S^{2}V^{3}
\end{equation*}
and the superconformal R charge by $\hat{r}=r+0.02q_{a}$. Let us comment on the two different operators of the form $S^4$ which are coupled to $F_6$ and $F_7$ in the superpotential. Denoting the representations under the nonabelian global symmetries by $(\boldsymbol{R}_{SU\left(2\right)_{s}},\boldsymbol{R}_{SU\left(3\right)_{v}})$, we have one gauge singlet of the form $S^2$ and it transforms as $(\textbf{1},\textbf{1})$, {\it i.e.} it is also a singlet of the global symmetries. Turning to operators of the form $S^4$, we have three independent gauge singlets: One corresponding to $(S^2)^2$ that transform as $(\textbf{1},\textbf{1})$, and another two that are not composed of smaller gauge-invariant operators and transform as $(\textbf{1},\textbf{1})$ and $(\textbf{5},\textbf{1})$. The later two operators are precisely the ones that are coupled to $F_6$ and $F_7$ in the superpotential.

Next, we examine the index at order $pq$ and find the following negative coefficient: 
\begin{equation}
-\boldsymbol{3}_{SU\left(2\right)_{s}}-\boldsymbol{5}_{SU\left(2\right)_{s}}-2\,\boldsymbol{8}_{SU\left(3\right)_{v}}-2=-\boldsymbol{8}_{SU\left(3\right)}-2\,\boldsymbol{8}_{SU\left(3\right)_{v}}-2\,.
\end{equation}
We see that the $SU\left(2\right)_{s}$ symmetry enhances to $SU(3)$ in the IR, and that instead of one $SU(3)_v$ current we observe two (due to the coefficient 2 in front of $\boldsymbol{8}_{SU\left(3\right)_{v}}$). This means that in the UV only one combination of these two symmetries is present, and we see it as $SU\left(3\right)_{v}$. In addition to these two enhancements, we also see an additional $U(1)$ symmetry in the IR. 

We conclude that the UV symmetry $SU\left(2\right)\times SU\left(3\right)\times U\left(1\right)$ enhances to $SU\left(3\right)^{3}\times U\left(1\right)^{2}$ in the IR, and that the dimension of the conformal manifold vanishes as there are no positive contributions (corresponding to marginal operators) at order $pq$.

As a consistency check, we can look as before at other parts of the index and check the enhancement. For example, denoting the $U(1)_a$ fugacity by $a$, we have at order $a^{12}(pq)^{1/2}$ the following coefficient: 
\begin{equation}
\boldsymbol{1}_{SU\left(2\right)_{s}}+\boldsymbol{3}_{SU\left(2\right)_{s}}+\boldsymbol{5}_{SU\left(2\right)_{s}}=\boldsymbol{3}_{SU\left(3\right)}+\boldsymbol{6}_{SU\left(3\right)}\,,
\end{equation}
and similar branching rules can be observed at other orders.

Like in the previous case, we have the complication of finding an additional $U(1)$ global symmetry at the IR. As a result, we again stress that the results found here are under the assumption that it does not mix with the superconformal R-symmetry. 

\

\subsection*{Spin(12)}

\

We consider the following matter content, 
\begin{center}
	\begin{tabular}{|c||c|c|c|c|c|}
		\hline
		Field & $Spin\left(12\right)_{g}$ & $SU\left(4\right)_{v}$ & $U\left(1\right)_{a}$ & $U\left(1\right)_{b}$ & $U\left(1\right)_{r}$\\
		\hline 
		
		$S$ & $\boldsymbol{32}'$ & $\boldsymbol{1}$ & 1 & 0 & $\frac{1}{4}$\\
		
		$C$ & $\boldsymbol{32}$ & $\boldsymbol{1}$ & 0 & 1 & $\frac{1}{4}$\\
		
		$V$ & $\boldsymbol{12}$ & $\boldsymbol{4}$ & -1 & -1 & 0\\
		
		$F_{0}$ & $\boldsymbol{1}$ & $\boldsymbol{6}$ & 0 & 2 & $\frac{3}{2}$\\
		
		$F_{1}$ & $\boldsymbol{1}$ & $\boldsymbol{6}$ & 2 & 0 & $\frac{3}{2}$\\
		
		$F_{2}$ & $\boldsymbol{1}$ & $\boldsymbol{\overline{10}}$ & 2 & 2 & 2\\
		
		$F_{3}$ & $\boldsymbol{1}$ & $\boldsymbol{\overline{4}}$ & 0 & 0 & $\frac{3}{2}$\\
		
		$F_{4}$ & $\boldsymbol{1}$ & $\boldsymbol{1}$ & -4 & 0 & 1\\
		
		$F_{5}$ & $\boldsymbol{1}$ & $\boldsymbol{1}$ & 0 & -4 & 1\\
		
		$F_{6}$ & $\boldsymbol{1}$ & $\boldsymbol{1}$ & -2 & -2 & 1\\
		
		$F_{7}$ & $\boldsymbol{1}$ & $\boldsymbol{\overline{4}}$ & -2 & 0 & 1\\
		
		$F_{8}$ & $\boldsymbol{1}$ & $\boldsymbol{\overline{4}}$ & 0 & -2 & 1\\
		
		$F_{9}$ & $\boldsymbol{1}$ & $\boldsymbol{4}$ & 2 & 2 & $\frac{3}{2}$\\
		\hline
	\end{tabular}
	\label{table:Spin12}
\end{center}
The superpotential is given by 
\begin{equation*}
W=F_{0}S^{2}V^{2}+F_{1}C^{2}V^{2}+F_{2}V^{2}+F_{3}SCV+F_{4}S^{4}+F_{5}C^{4}+F_{6}S^{2}C^{2}+F_{7}S^{3}CV+F_{8}SC^{3}V+F_{9}SCV^{3}
\end{equation*}
and the superconformal R charge by $\hat{r}=r-0.123\left(q_{a}+q_{b}\right)$. Calculating the index, we find the following expansion: 
\begin{equation*}
\mathcal{I}=1+\left[\boldsymbol{6}_{SU\left(4\right)_{v}}+\boldsymbol{10}_{SU\left(4\right)_{v}}+a^{-4}+b^{-4}+2a^{-2}b^{-2}+2\,\boldsymbol{\overline{4}}_{SU\left(4\right)_{v}}\left(a^{-2}+b^{-2}\right)\right]\left(pq\right)^{\frac{1}{2}}+
\end{equation*}
\begin{equation*}
+2\left[\boldsymbol{6}_{SU\left(4\right)_{v}}\left(a^{2}+b^{2}\right)+\boldsymbol{\overline{4}}_{SU\left(4\right)_{v}}+\boldsymbol{4}_{SU\left(4\right)_{v}}a^{2}b^{2}\right]\left(pq\right)^{\frac{3}{4}}+
\end{equation*}
\begin{equation*}
+\left[a^{2}b^{2}\left(\boldsymbol{6}_{SU\left(4\right)_{v}}+\boldsymbol{\overline{10}}_{SU\left(4\right)_{v}}\right)+2\,\boldsymbol{20'}_{SU\left(4\right)_{v}}+\boldsymbol{35}_{SU\left(4\right)_{v}}+\boldsymbol{45}_{SU\left(4\right)_{v}}-\boldsymbol{15}_{SU\left(4\right)_{v}}-2\right]pq+\ldots
\end{equation*}
where $a$ and $b$ are the fugacities of $U(1)_a$ and $U(1)_b$, respectively. We notice that upon changing the basis of these two $U(1)$ symmetries as follows,
\begin{equation}
U\left(1\right)_{e}=\frac{1}{2}\left[U\left(1\right)_{a}-U\left(1\right)_{b}\right]\,,\,\,\,\,U\left(1\right)_{h}=\frac{1}{2}\left[U\left(1\right)_{a}+U\left(1\right)_{b}\right]\,,
\end{equation}
only $U(1)_h$ mixes with $U(1)_r$ in the expression for the superconformal R charge, and the fugacity of $U(1)_e$ forms characters of $SU(2)$. More explicitly, the index can be written using representations of $SU(2)$ as follows: 
\begin{equation*}
\mathcal{I}=1+\left[\boldsymbol{6}_{SU\left(4\right)_{v}}+\boldsymbol{10}_{SU\left(4\right)_{v}}+\left(1+\boldsymbol{3}_{SU\left(2\right)}\right)h^{-2}+2\,\boldsymbol{\overline{4}}_{SU\left(4\right)_{v}}\boldsymbol{2}_{SU\left(2\right)}h^{-1}\right]\left(pq\right)^{\frac{1}{2}}+
\end{equation*}
\begin{equation*}
+2\left[\boldsymbol{6}_{SU\left(4\right)_{v}}\boldsymbol{2}_{SU\left(2\right)}h+\boldsymbol{\overline{4}}_{SU\left(4\right)_{v}}+\boldsymbol{4}_{SU\left(4\right)_{v}}h^{2}\right]\left(pq\right)^{\frac{3}{4}}+
\end{equation*}
\begin{equation*}
+\left[h^{2}\left(\boldsymbol{6}_{SU\left(4\right)_{v}}+\boldsymbol{\overline{10}}_{SU\left(4\right)_{v}}\right)+2\,\boldsymbol{20'}_{SU\left(4\right)_{v}}+\boldsymbol{35}_{SU\left(4\right)_{v}}+\boldsymbol{45}_{SU\left(4\right)_{v}}-\boldsymbol{15}_{SU\left(4\right)_{v}}-2\right]pq+\ldots
\end{equation*}
where $h$ is the fugacity of $U(1)_h$, and the superconformal R charge is given by $\hat{r}=r-0.246q_{h}$. This suggests that $U(1)_e$ enhances to $SU(2)$ somewhere on the conformal manifold, in which case there is a marginal operator at this point that transforms in the adjoint of $SU(2)$ such that its contribution to the index at order $pq$ cancels that of the $SU(2)$ current. 

Let us examine the coefficient at order $pq$ (with a vanishing $U(1)_h$ charge) more closely. From the expansion above, it is given by: 
\begin{equation}
\label{cpq}
C_{pq}=2\,\boldsymbol{20'}_{SU\left(4\right)_{v}}+\boldsymbol{35}_{SU\left(4\right)_{v}}+\boldsymbol{45}_{SU\left(4\right)_{v}}-\boldsymbol{15}_{SU\left(4\right)_{v}}-2
\end{equation}
and we would like to match it with the general form \eqref{indform}:
\begin{equation}
\label{cpqgen}
C_{pq}=\textrm{Marginals-Currents}\,.
\end{equation}
At points in the conformal manifold where there is no enhancement of the symmetry, we clearly have 
\begin{equation}
\textrm{Marginals}=2\,\boldsymbol{20'}_{SU\left(4\right)_{v}}+\boldsymbol{35}_{SU\left(4\right)_{v}}+\boldsymbol{45}_{SU\left(4\right)_{v}}
\end{equation}
and 
\begin{equation}
\textrm{Currents}=\boldsymbol{15}_{SU\left(4\right)_{v}}+2\,,
\end{equation}
corresponding to the UV symmetry $SU\left(4\right)_{v}\times U\left(1\right)_{a}\times U\left(1\right)_{b}$. Now, since we have two relevant operators of the form $S^{2}C^{2}V^{2}$ that have superconformal R charge 1 and transform in the $\boldsymbol{6}_{SU\left(4\right)_{v}}$ and in the $\boldsymbol{10}_{SU\left(4\right)_{v}}$ (see the first two terms in the expansion above at order $(pq)^{1/2}$), it is natural to expect that there is a point in the conformal manifold where they give rise to marginal operators of the form $\left(S^{2}C^{2}V^{2}\right)^{2}$ that transform in the representations: 
\begin{equation}
\label{margen1}
\textrm{Marginals}=1+\boldsymbol{15}_{SU\left(4\right)_{v}}+2\,\boldsymbol{20'}_{SU\left(4\right)_{v}}+\boldsymbol{35}_{SU\left(4\right)_{v}}+\boldsymbol{45}_{SU\left(4\right)_{v}}\,.
\end{equation}
From \eqref{cpq} and \eqref{cpqgen}, this means that at such a point: 
\begin{equation}
\label{curren1}
\textrm{Currents}=2\,\boldsymbol{15}_{SU\left(4\right)_{v}}+3\,,
\end{equation}
corresponding to the larger symmetry group $SU\left(4\right)^{2}\times U\left(1\right)^{3}$. 

Recall that the expansion of the index suggests that there is also a point where one of the $U(1)$ groups enhances to $SU(2)$. If there is a point at which the two kinds of enhancements take place, it means that at this point two more marginal operators that contribute $e^2+e^{-2}$ to the index at order $pq$ add to \eqref{margen1}, along with two more conserved currents with the same contribution but with an opposite (minus) sign that add to \eqref{curren1}. This way, the adjoint representation of $SU(2)$ ($\boldsymbol{3}_{SU\left(2\right)}=1+e^{2}+e^{-2}$) is obtained and we have: 
\begin{equation}
\textrm{Marginals}=\boldsymbol{3}_{SU\left(2\right)}+\boldsymbol{15}_{SU\left(4\right)_{v}}+2\,\boldsymbol{20'}_{SU\left(4\right)_{v}}+\boldsymbol{35}_{SU\left(4\right)_{v}}+\boldsymbol{45}_{SU\left(4\right)_{v}}
\end{equation}
and 
\begin{equation}
\textrm{Currents}=2\,\boldsymbol{15}_{SU\left(4\right)_{v}}+\boldsymbol{3}_{SU\left(2\right)}+2\,.
\end{equation}
Since the representations and operators that appear in the index are consistent with this enhancement, we conjecture that such a point indeed exists, at which the UV symmetry $SU\left(4\right)\times U\left(1\right)^{2}$ enhances to $SU\left(4\right)^{2}\times SU\left(2\right)\times U\left(1\right)^{2}$.

Note that the analysis of this model, including the various arguments, is reminiscent of the analysis of the $Spin(8)$ model performed in \cite{Razamat:2018gbu}. Note also that both of these models correspond to the $n=4$ case of the two $Spin$ sequences -- $Spin(n+8)$ and $Spin(n+4)$ -- discussed here and in \cite{Razamat:2018gbu}. 

We again note that these results suggest the existence of an additional $U(1)$ symmetry, and so we must further assume that it does not mix with the superconformal R-symmetry. 

\

\subsection*{Spin(13)}

\

We consider the model given in the table below. 
\begin{center}
	\begin{tabular}{|c||c|c|c|c|}
		\hline
		Field & $Spin\left(13\right)_{g}$ & $SU\left(5\right)_{v}$ & $U\left(1\right)_{a}$ & $U\left(1\right)_{r}$\\
		\hline 
		
		$S$ & $\boldsymbol{64}$ & $\boldsymbol{1}$ & 5 & $\frac{1}{4}$\\
		
		$V$ & $\boldsymbol{13}$ & $\boldsymbol{5}$ & -8 & 0\\
		
		$F_{0}$ & $\boldsymbol{1}$ & $\boldsymbol{\overline{10}}$ & 6 & $\frac{3}{2}$\\
		
		$F_{1}$ & $\boldsymbol{1}$ & $\boldsymbol{10}$ & 14 & $\frac{3}{2}$\\
		
		$F_{2}$ & $\boldsymbol{1}$ & $\boldsymbol{\overline{5}}$ & -12 & 1\\
		
		$F_{3}$ & $\boldsymbol{1}$ & $\boldsymbol{\overline{15}}$ & -4 & 1\\
		
		$F_{4}$ & $\boldsymbol{1}$ & $\boldsymbol{\overline{15}}$ & 16 & 2\\
		
		$F_{5}$ & $\boldsymbol{1}$ & $\boldsymbol{1}$ & -20 & 1\\
		\hline
	\end{tabular}
	\label{table:Spin13}
\end{center}
The superpotential is given by 
\begin{equation*}
W=F_{0}S^{2}V^{2}+F_{1}S^{2}V^{3}+F_{2}S^{4}V+F_{3}S^{4}V^{2}+F_{4}V^{2}+F_{5}S^{4}
\end{equation*}
and the superconformal R charge by $\hat{r}=r-0.0289q_{a}$. All the gauge-invariant operators are above the unitarity bound and the superpotential is a relevant deformation of the theory defined without it. 

Calculating the coefficient at order $pq$ in the expansion of the index, we find: 
\begin{equation}
-2\,\boldsymbol{24}_{SU(5)_v}-2\,,
\end{equation}
suggesting (under the usual assumptions) that the UV symmetry $SU\left(5\right)\times U\left(1\right)$ enhances to $SU\left(5\right)^{2}\times U\left(1\right)^{2}$ in the IR. Moreover, we see that the dimension of the conformal manifold vanishes as there are no positive contributions at order $pq$ corresponding to marginal operators, and that the UV symmetry is doubled at this IR fixed point. 

\

\subsection*{Spin(14)}

\

The last model in the sequence corresponds to $n=6$, and as mentioned in the beginning of this section we consider here two (in fact, three) models, one with an enhanced symmetry that fits the pattern we observe and another with an even larger enhancement. We begin with a discussion of the model with the larger enhancement, and then show how to obtain from it two different models with an enhancement that fits our pattern. All the models discussed in this subsection are closely related and only differ by the choice of superpotential.

We begin with the model given in the table below. 
\begin{center}
	\begin{tabular}{|c||c|c|c|c|}
		\hline
		Field & $Spin\left(14\right)_{g}$ & $SU\left(6\right)_{v}$ & $U\left(1\right)_{a}$ & $U\left(1\right)_{r}$\\
		\hline 
		
		$S$ & $\boldsymbol{\overline{64}}$ & $\boldsymbol{1}$ & 3 & $\frac{1}{4}$\\
		
		$V$ & $\boldsymbol{14}$ & $\boldsymbol{6}$ & -4 & 0\\
		
		$F_{0}$ & $\boldsymbol{1}$ & $\boldsymbol{20}$ & 6 & $\frac{3}{2}$\\
		
		$F_{1}$ & $\boldsymbol{1}$ & $\boldsymbol{\overline{21}}$ & -4 & 1\\
		
		$F_{2}$ & $\boldsymbol{1}$ & $\boldsymbol{\overline{21}}$ & 8 & 2\\
		
		$F_{3}$ & $\boldsymbol{1}$ & $\boldsymbol{1}$ & -24 & 0\\
		\hline
	\end{tabular}
	\label{table:Spin14}
\end{center}
The superpotential is given by 
\begin{equation}
\label{W14e}
W=F_{0}S^{2}V^{3}+F_{1}S^{4}V^{2}+F_{2}V^{2}+F_{3}S^{8}
\end{equation}
and the superconformal R charge by $\hat{r}=r-0.0528q_{a}$. As in all the models in this sequence, all the gauge-invariant operators are above the unitarity bound and the superpotential is a relevant deformation of the theory defined without it. 

Calculating the index at order $pq$, we find: 
\begin{equation}
\label{en141}
-3\,\boldsymbol{35}_{SU\left(6\right)_{v}}\,.
\end{equation}
Now, since we know that the symmetry $U(1)_a$ is present also in the IR, we conclude that there is a marginal operators in the singlet of $SU(6)_v$ that cancels the contribution of the $U(1)_a$ current [recall the general form \eqref{indform}]. Moreover, as we do not see any evidence for an even larger symmetry, we conjecture from \eqref{en141} that the conformal manifold is one dimensional and on every point of which, the UV symmetry $SU\left(6\right)\times U\left(1\right)$ enhances to $SU\left(6\right)^{3}\times U\left(1\right)$.

Let us examine this enhancement more closely, and check where the two extra copies of $SU\left(6\right)$ come from. Following the discussion in section 5 of \cite{Razamat:2018gbu}, we first consider the model obtained from the one above after deleting the first two terms in the superpotential \eqref{W14e}. That is, we do not flip the operators $S^{2}V^{3}$ and $S^{4}V^{2}$ (in the language we introduced before) and remain with the superpotential 
\begin{equation}
W=F_{2}V^{2}+F_{3}S^{8}\,.
\end{equation}
In this model, the operators $S^{2}V^{3}$ and $S^{4}V^{2}$ are not set to zero in the chiral ring, and from analyzing the various representations under the global symmetry we find the following marginal relations: 
\begin{equation}
\left.\left(S^{2}V^{3}\right)\left(S^{6}V^{3}\right)\right|_{\boldsymbol{35}_{SU\left(6\right)_{v}}}\sim0
\end{equation}
and
\begin{equation}
\left.\left(S^{4}V^{2}\right)\left(S^{4}V^{4}\right)\right|_{\boldsymbol{35}_{SU\left(6\right)_{v}}}\sim0\,.
\end{equation}
These relations result from the compositeness of the smaller operators, and state that the representation $\boldsymbol{35}_{SU\left(6\right)_{v}}$ is missing from the marginal operators obtained by multiplying them. In fact, among all the (marginal) operators that are built from 8 spinors and 6 vectors, only $\left(S^{8}V^{4}\right)\left(V^{2}\right)$ corresponds to an operator that transforms in the $\boldsymbol{35}_{SU\left(6\right)_{v}}$. However, since $V^2$ is flipped by $F_2$, this operator does not appear in the spectrum of the IR theory. 

Next, as shown in \cite{Razamat:2018gbu}, we can obtain models with extra IR $SU(6)$ symmetries by flipping either $S^{2}V^{3}$, $S^{4}V^{2}$ or both of them. Each such single flip would yield a model with one extra $SU(6)$, and flipping both of the operators as in the model we analyzed above corresponds to two extra $SU(6)$ symmetries. Therefore, in the two models corresponding to the superpotentials   
\begin{equation}
W=F_{0}S^{2}V^{3}+F_{2}V^{2}+F_{3}S^{8}
\end{equation}
and 
\begin{equation}
\label{W14p}
W=F_{1}S^{4}V^{2}+F_{2}V^{2}+F_{3}S^{8}\,,
\end{equation}
we conjecture that there is a point on their conformal manifolds at which the UV symmetry $SU\left(6\right)\times U\left(1\right)$ enhances to $SU\left(6\right)^{2}\times U\left(1\right)$. Both of these models present the same enhancement, and one of them is expected to correspond to our sequence. Notice, however, that also the second one might have a six-dimensional lift. 

\

\subsection*{Summary of the symmetry enhancements}

\

The UV and IR symmetries we obtained in the various models are summarized in the table below. 
\begin{center}
	\begin{tabular}{|c|c|c|c|c|}
		\hline
		Gauge group & UV global symmetry & IR global symmetry & Rank at UV & Rank at IR\\
		\hline 
		\hline 
		$Spin\left(9\right)$ & $SU\left(4\right)\times U\left(1\right)$ & $SU\left(6\right)\times U\left(1\right)$ & 4 & 6\\
		\hline 
		$Spin\left(10\right)$ & $SU\left(2\right)^{3}\times U\left(1\right)^{2}$ & $SU\left(4\right)\times SU\left(2\right)\times U\left(1\right)^{3}$ & 5 & 7\\
		\hline 
		$Spin\left(11\right)$ & $SU\left(2\right)\times SU\left(3\right)\times U\left(1\right)$ & $SU\left(3\right)^{3}\times U\left(1\right)^{2}$ & 4 & 8\\
		\hline 
		$Spin\left(12\right)$ & $SU\left(4\right)\times U\left(1\right)^{2}$ & $SU\left(4\right)^{2}\times SU\left(2\right)\times U\left(1\right)^{2}$ & 5 & 9\\
		\hline 
		$Spin\left(13\right)$ & $SU\left(5\right)\times U\left(1\right)$ & $SU\left(5\right)^{2}\times U\left(1\right)^{2}$ & 5 & 10\\
		\hline 
		$Spin\left(14\right)$ & $SU\left(6\right)\times U\left(1\right)$ & $SU\left(6\right)^{2}\times U\left(1\right)$ & 6 & 11\\
		\hline
	\end{tabular}
	\label{table:Sammary}
\end{center}
The following sequence of groups: 
\begin{equation}
SU\left(6\right),\,\,SU\left(4\right)\times U\left(1\right),\,\,SU\left(3\right)\times U\left(1\right),\,\,SU\left(2\right)\times U\left(1\right),\,\,U\left(1\right),\,\,\phi
\end{equation}
is exactly given by the commutant of $SU\left(n\right)\times SU\left(2\right)\times SU\left(3\right)$ in $E_8$, and we can therefore identify the following pattern of IR symmetries: 
\begin{equation}
\textrm{4d IR symmetry:}\,\,\,\,\,\,\,C_{E_{8}}\left(SU\left(n\right)\times SU\left(2\right)\times SU\left(3\right)\right)\times SU\left(n\right)^{2}\times U\left(1\right)
\end{equation}
or 
\begin{equation}
\label{4dpatt8}
\textrm{4d IR symmetry:}\,\,\,\,\,\,\,C_{E_{9-n}}\left(SU\left(2\right)\times SU\left(3\right)\right)\times SU\left(n\right)^{2}\times U\left(1\right),
\end{equation}
where we use the notations of subsection \ref{np4}, and the only exception is $n=2$ where one of the $SU(2)$ groups is broken to a $U(1)$. Note the close resemblance with the result \eqref{4dpatt} of \cite{Razamat:2018gbu}, and in particular the breaking of one of the $SU(2)$ groups to $U(1)$ in the $n=2$ case. We see that the expressions \eqref{4dpatt8} and \eqref{4dpatt} only differ by an additional $SU(3)$ in the commutant, which might correspond in the six dimensional construction to an $SU(3)$ gauging of the $6d$ SCFTs we considered before (see subsection \ref{np4}). Let us then turn to discuss the compactification from $6d$.

%% file: sections/sixdimnew.tex
\section{Compactification from six dimensions}
\label{sixdim}

In this section we shall conjecture an explanation for the observed symmetry enhancement. The explanation will be in terms of the compactification of $6d$ SCFTs on a torus with fluxes. In order to understand the logic of the approach we first review the methods introduced in \cite{Kim:2017toz,Kim:2018bpg,Kim:2018lfo} to conjecture $4d$ field theories generated by such compactifications. We start with a $6d$ $\mathcal{N}=(1,0)$ SCFT that we wish to compactify on a torus to four dimensions with fluxes in its global symmetry supported on the torus. The approach generally used to tackle this problem is to compactify first on one circle to $5d$, study the resulting theory and then continue with the compactification to $4d$. The advantage of compactifying first to $5d$ is that it allows us to exploit a relation between $6d$ SCFTs and $5d$ gauge theories.

The specific relation that we seek to exploit is that when compactified on a circle to $5d$, potentially with an holonomy in the global symmetry, the $6d$ SCFT can flow at low-energies to a $5d$ gauge theory. The most well known examples of this are the $\mathcal{N}=(2,0)$ $6d$ SCFTs that reduce to $5d$ maximally supersymmetric Yang-Mills theories \cite{Douglas:2010iu,Lambert:2010iw}. These types of relations are ubiquitous also for $(1,0)$ SCFTs. For instance, the rank $N$ E-string $6d$ SCFTs, compactified on a circle, are known to reduce to the $5d$ $USp(2N)$ gauge theories with an antisymmetric hypermultiplet and eight fundamental hypermultiplets, assuming a suitable holonomy is turned on \cite{Ganor:1996pc}. Similar relations have also been observed for many other $6d$ SCFTs and $5d$ gauge theories, see for instance \cite{Hayashi:2015fsa,Zafrir:2015rga,Hayashi:2015zka,Hayashi:2015vhy}.

Now let us return to the problem at hand, the compactification of a $6d$ $\mathcal{N}=(1,0)$ SCFT on a torus with fluxes in its global symmetry supported on the torus. We would like to first compactify on one circle to $5d$ so that the $6d$ SCFT reduces to a $5d$ gauge theory. However, we need to consider the effect of the flux. This was analyzed in \cite{Kim:2017toz,Kim:2018bpg,Kim:2018lfo}, and the conclusion reached there is that the flux leads to domain walls between different $5d$ gauge theories. The resulting picture is that when we reduce on the first circle to $5d$ we end up with multiple copies of the $5d$ gauge theory description of the $6d$ SCFT interacting through the fields living on the domain walls. Unlike the bulk, the domain walls preserve only half of the supersymmetry, corresponding to $\mathcal{N}=1$ in four dimensions. This corresponds to the fact that the flux preserves only $\mathcal{N}=1$ supersymmetry in $4d$ \cite{Razamat:2016dpl}. Generically the bulk matter is subjected to various $\mathcal{N}=1$ (only) preserving boundary conditions on the domain walls, and these usually reduce the $4d$ $\mathcal{N}=2$ vector and hypermultiplets that are expected to come from the bulk $\mathcal{N}=1$ $5d$ fields to $4d$ $\mathcal{N}=1$ vector and chiral fields, respectively.

We can now reduce along the second circle to $4d$. The $5d$ bulk is straightforward to reduce, as it consists only of IR free gauge theories, and reduces to the same gauge theories in $4d$, though some of the $5d$ bulk matter is killed by Dirichlet boundary conditions on the domain walls. These then interact with one another via the matter living on the domain walls. This chain of thought suggests that the $4d$ theories we get from the compactification of a $6d$ SCFT on a torus with fluxes will resemble a circle of multiple copies of a $4d$ $\mathcal{N}=1$ version of the $5d$ gauge theory. By $\mathcal{N}=1$ version we mean the same gauge theory, but with the $4d$ $\mathcal{N}=2$ vector and hypermultiplets, that are the usual analogues of $5d$ $\mathcal{N}=1$ matter, being replaced by $4d$ $\mathcal{N}=1$ vector and chiral fields. The number of copies is related to the number of domain walls which in turn is related to the value of the flux. In the minimal flux case, we expect just one copy.

We are now ready to discuss the way we shall seek an explanation for the $4d$ symmetry enhancement. For this, we consider $5d$ gauge theories with matter content similar to that of the $4d$ theories we considered, but with $4d$ $\mathcal{N}=1$ vector and chiral fields replaced with $5d$ $\mathcal{N}=1$ vector fields and hypermultiplets. We also consider $6d$ SCFTs that upon circle compactification reduce to these $5d$ gauge theories. Then on one hand, by the preceding picture, it is reasonable that we may be able to get the $4d$ gauge theories we studied in the previous section by the torus compactification with fluxes of the $6d$ SCFTs. On the other hand, the IR $4d$ theory resulting from the compactification should inherit the global symmetry of the $6d$ SCFT, up to the breaking incurred by the flux. Stated in other words, a $6d$ SCFT on a compact manifold with fluxes can be viewed at low enough energies as a $4d$ SCFT with a global symmetry given by the subgroup of the original $6d$ symmetry that commutes with the fluxes. If this symmetry is greater than that visible in the UV $4d$ gauge theory then an enhancement is expected in the IR to conform with the $6d$ expectations. This is the nature of the explanation for the enhancement that we shall propose.

We next turn to analyze this in greater detail. We shall first consider the $5d$ variants of the $4d$ theories we studied in the previous section. Using known results, we argue that they should indeed be low-energy descriptions of $6d$ SCFTs on a circle, and propose a conjecture for the identity of the $6d$ SCFTs in question. We shall then also argue that the global symmetry of these $6d$ SCFTs is such that our proposed explanation can work. We then close the section with describing various aspects of the reduction itself. 

\          

\subsection{Five dimensional analysis}

Here we consider the family of five dimensional $\mathcal{N}=1$ gauge theories given by a $Spin(n+8)$ gauge group, $n$ hypermultiplets in the vector representation and a total of $64$ components of hypermultiplets that are in the spinor representation. This is the $5d$ $\mathcal{N}=1$ version of the $4d$ gauge theories studied in the previous section, modulo gauge singlets. The first thing we note about this family is that it is expected to be a low-energy description of a $6d$ $\mathcal{N}=(1,0)$ SCFT on a circle. To understand how this comes about, it is useful to first recall some aspects of this type of relations. For this, we shall use the example of the $6d$ $\mathcal{N}=(2,0)$ SCFT and $5d$ maximally supersymmetric Yang-Mills theory. The claim here is that the latter still contains some information about the massive Kaluza-Klein excitations. Notably, the $5d$ gauge theory possesses massive non-perturbative excitations given by its instanton particle, and for the case at hand these are supposed to contribute the additional Kaluza-Klein modes.

One way to see that a $5d$ gauge theory may be the low-energy description of a $6d$ $\mathcal{N}=(1,0)$ SCFT on a circle is to find the existence of such modes, particularly, the ones associated with conserved currents. In the $5d$ theory the Kaluza-Klein modes of the $6d$ conserved currents are given by broken conserved current multiplets. This is as the currents are expected to be conserved when the radius of the compactification is taken to zero and we get the $6d$ SCFT on a flat spacetime. In this type of relations, the radius of compactification is inversely related to the coupling constant of the $5d$ gauge theory so the zero radius limit maps to the infinite coupling limit of the gauge theory. Therefore, from the $5d$ gauge theory viewpoint, the only thing breaking the currents is a mass deformation associated with the coupling constant, which has positive mass dimensions in $5d$. This is similar to the phenomena of enhancement of symmetry in $5d$ gauge theories, originally discovered in \cite{Seiberg:1996bd}. In that case the $5d$ gauge theory is a low-energy description of a $5d$ SCFT deformed by a mass deformation. The mass deformation can break the global symmetry whose broken currents manifest in the low-energy gauge theory as instanton particles. The major difference between the two cases is that while in the latter case the low-energy global symmetry plus the instantonic broken currents form a finite Lie group, the global symmetry group of the $5d$ SCFT, in the former case it instead forms an affine Lie group. This affine Lie group essentially describes all the Kaluza-Klein modes of the $6d$ conserved current multiplet. As a result, the appearance of instantonic broken currents in a $5d$ gauge theory whose spectrum is such that they form an affine Lie group is generally an indication that this $5d$ gauge theory is a low-energy description of a $6d$ SCFT on a circle.

In \cite{Tachikawa:2015mha} a method to study the instantonic broken currents provided by $1$-instanton configurations was devised and in \cite{Zafrir:2015uaa} it was used to study cases that include the family of theories that is of interest to us here. It was found there that the resulting spectrum is not consistent with a finite Lie group, but is consistent with an affine one. This is a first indication that this class of theories lift to $6d$ SCFTs, although the analysis there only takes into account broken currents coming from $1$-instanton configurations and so does not see contribution from higher order instantons which could spoil this expected behavior. Nevertheless, ultimately, the question of when does a $5d$ gauge theory possess a UV completion as a $5d$ or $6d$ SCFT is currently unanswered so we cannot say for sure if these gauge theories lift to $6d$ SCFTs or not. However, \cite{Jefferson:2017ahm} has put forward several criteria in an attempt to answer this question. These criteria are known to be insufficient though they do appear to be necessary. The family of theories we study here all fit the criteria for a $6d$ lifting theory. As a result there are several indications that this family lifts to $6d$ SCFTs though it is not assured\footnote{Recently, there has been some progress in the study of $6d$ SCFT lifts of $5d$ gauge theories using geometric methods \cite{ZHM,EJeK,EJaK1,EKY,Jefferson:2018smsm,EK1,EK2,BJ1,BJ2,ALM,EJaK2,EJ,ALLSW1,ALLSW2,ALLSW3,Bhardwaj:2019cls,BJKTV}. It is our hope that these tools may also be useful for the cases discussed here.}.

\subsubsection*{Determining the $6d$ lifts}

In what follows we shall assume that this family indeed lifts to $6d$ SCFTs, and try to determine the $6d$ SCFTs it can lift to. Finding a known family of $6d$ $\mathcal{N}=(1,0)$ SCFTs that has the correct properties to reduce to these $5d$ gauge theories, is in itself a non-trivial test that these $5d$ gauge theories are $6d$ lifting. To do this we need to consider what properties do we expect from the $6d$ SCFTs. There are several consistency conditions that should be obeyed. First, the global symmetry of the $6d$ SCFTs must be consistent with that expected from the $5d$ gauge theories when the instantonic broken currents are included. Here we only have partial information as we only know the ones provided by the $1$-instanton. A second requirement is that the Coulomb branch dimension of the $5d$ gauge theory be correctly reproduced by the $6d$ SCFT. Generally, when reducing a $6d$ SCFT on a circle, the Coulomb branch of the resulting $5d$ theory receives contributions from two sources. One is the $6d$ vector multiplets on a circle and the other is the tensor multiplets that are dual to vector multiplets in $5d$. As a result the $5d$ Coulomb branch dimension expected for the low-energy $6d$ SCFT on a circle is given by the sum of the dimension of the $6d$ tensor branch and the total rank of all the gauge groups of the low-energy gauge theory on a generic point on the tensor branch. 

The final test we can perform is to consider the behavior under deformations, notably Higgs branch flows. The Higgs branch has the interesting property of being invariant under a dimensional reduction, and as a result Higgs branch flows done on the $6d$ SCFTs should lead to analogous flows on the $5d$ gauge theories, and vice versa. This can be used as an additional test by flowing from the $5d$ gauge theories to other theories whose lift is known, and seeing if we get the expected $6d$ SCFTs by an analogous flow on the conjectured $6d$ SCFT lifts. It should be noted that while the Higgs branch is invariant under a dimensional reduction, it does change under deformations such as a mass deformation, Coulomb branch or tensor branch vevs, and so the Higgs branch of the SCFT and that of its associated $5d$ or $6d$ gauge theory description may differ (see for instance \cite{Cremonesi:2015lsa,Ferlito:2017xdq,Hanany:2018uhm,HaZ}). Ultimately, the analysis we perform involves some form of deformation of this type and so we do not actually see the entire Higgs branch, which may limit the use of this method.

Next we shall study the global symmetry expected from the $6d$ SCFT based on the $5d$ gauge theory, and use it to conjecture the $6d$ SCFT lift. We then subject that conjecture to various consistency checks. As previously mentioned the global symmetry at the UV, expected from both the classically visible one and the contribution of $1$-instanton states, was studied in \cite{Zafrir:2015uaa}. We summarize the results found there in table \ref{table:gs}. In the table we have also included two other cases involving the $Spin(12)$ gauge group, where the case symmetric under the exchange of the two types of spinors is the one that we naturally associate with the family. This comes about as there is a sequence of Higgs branch flows connecting the theories. Starting from the $Spin(14)$ case, the $Spin(12)$ case we get is the symmetric one. We shall refer to this line, starting with the $Spin(14)$ case, as the main line. The other cases then lie outside it, each starting a new line that joins the main one at $Spin(11)$. We also note that for Lie groups with complex representations we get the twisted affine groups. This implies that the reduction here should be done with a twist that acts like the complex conjugation outer automorphism on these groups.

From these results it is possible to infer the global symmetry of the expected $6d$ SCFT. Specifically, we expect the $6d$ SCFT to have the finite group associated with the affine Lie group. In cases where the group is not affinized at the $1$-instanton level, we assume that this happens at a higher instanton order and we get the same finite Lie group in $6d$. Taking the $5d$ gauge group to be $Spin(n+8)$, we note that the global symmetry of the main line, expected of the $6d$ SCFTs, can be written concisely as $C_{E_{9-n}}\left(SU\left(3\right)\right)\times SU\left(2n\right)$, with the exception of the $n=2$ case where the $SU(4)$ is enhanced to $Spin(7)$.   

\begin{table}[h!]
\begin{center}
\begin{tabular}{|c|c|}
  \hline 
  Theory  & Symmetry  \\
\hline
   $Spin(9)+1V+4S$ & $E^{(2)}_{6}\times SU(2)$ \\ 
\hline
   $Spin(10)+2V+4S$ & $B^{(1)}_{3}\times A^{(2)}_{5}$ \\ 
\hline
   $Spin(11)+3V+2S$ & $A^{(2)}_{5}\times A^{(2)}_{2}\times A^{(2)}_{2}$ \\  
\hline
   $Spin(12)+4V+1S+1C$ & $A^{(2)}_{7}\times A^{(1)}_{1}\times A^{(1)}_{1}\times U(1)^{(2)}$ \\ 
\hline
   $Spin(12)+4V+\frac{3}{2}S+\frac{1}{2}C$ & $A^{(2)}_{2}\times USp(8)$ \\ 
\hline
   $Spin(12)+4V+2S$ & $E^{(2)}_{6}\times A^{(2)}_{2}\times A^{(2)}_{2}$ \\ 
\hline
   $Spin(13)+5V+1S$ & $A^{(2)}_{9}\times A^{(1)}_{1}\times U(1)^{(2)}$ \\
\hline
   $Spin(14)+6V+1S$ & $A^{(2)}_{11}\times A^{(1)}_{1}$ \\ 
\hline
\end{tabular}
 \end{center}
\label{table:gs}
\caption{The minimal symmetry consistent with the perturbative plus $1$-instanton contribution for the $5d$ $Spin$ gauge theories considered here. The spectrum of the expected additional current is such that it can only be accommodated by an affine group, at least for some factors of the global symmetry. This is interpreted as the theory lifting to a $6d$ SCFT in the infinite coupling limit, whose symmetry is the finite Lie group associated with the affine case. The superscript here denotes whether the affine group is the twisted ($2$) or the untwisted version ($1$). This lifts to whether the compactification of the $6d$ SCFT involves a twist or not. We also use the notation of $U(1)^{(2)}$ for a $U(1)$ group projected out by charge conjugation.}
\end{table}        

This enables us to formulate a conjecture for the identity of the $6d$ SCFTs in question. Specifically, we conjecture that the $6d$ SCFT lifts have tensor branch descriptions given by a gauging of the rank $1$ E-string theory by a pure $SU(3)$ gauge group and an additional $SU(n)$ gauge group with $2n$ hypermultiplets in the fundamental representation. This combination of matter gives an anomaly free theory, once tensor multiplets are introduced for the two gauge groups and used to cancel their respective gauge anomalies. Therefore, we expect it to be a low-energy description of a $6d$ SCFT, deformed by going on the tensor branch associated with the coupling constants of the two gauge groups. This $6d$ SCFT is our conjectured UV completion of the $Spin(n+8)$ family of theories.

This conjecture indeed reproduces the desired global symmetry as we get the $SU\left(2n\right)$ from the symmetry rotating the $2n$ fundamental hypermultiplets, and $C_{E_{9-n}}\left(SU\left(3\right)\right)$ is by definition the commutant of $SU(3)\times SU(n)$ inside $E_8$. Here the $U(1)$ baryon symmetry naively expected from the fundamental hypermultiplets is anomalous and so is not actually a global symmetry. The only exceptional case is the $n=2$ case as now we have an $SU(2)$ gauge theory with four fundamental hypermultiplets. Because the fundamental representation of $SU(2)$ is self-conjugate, the $SU(4)$ symmetry we naively expect to rotate four complex hypermultiplets is enhanced. Naively we expect it to enhance to $SO(8)$, but as discussed in \cite{Ohmori:2015pia}, it turns out that the SCFT only exhibits a $Spin(7)$ symmetry. Therefore, in this case we expect the global symmetry to be $Spin(7)\times SU(6)$, in accordance with the symmetry expected from the associated $5d$ gauge theory. 

Another noteworthy case is the $n=1$ case. Here we naively have an $SU(1)$ gauging of the rank $1$ E-string theory. Here the $SU(1)$ part is really just the tensor multiplet that always accompanies any gauge group in low-energy gauge theories related to $6d$ SCFTs on the tensor branch. This tensor joins with the rank $1$ E-string tensor to give the rank $2$ E-string SCFT. As a result, this case is better described as an $SU(3)$ gauging of the rank $2$ E-string SCFT. This is indeed an anomaly free theory, once a tensor multiplet is introduced for the $SU(3)$ gauge group and used to cancel the $SU(3)$ gauge anomalies, and should descend from a $6d$ SCFT deformed by a vev to the scalar in the tensor multiplet.   

\subsubsection*{Evidence for the conjecture}

After we saw that our conjecture correctly reproduces the expected global symmetry we next consider other evidence for it. First we note that the end point of the main line is the same for both the $5d$ and $6d$ theories. Notably, we cannot have $64$ spinor degrees of freedom for $Spin(n+8)$ if $n>6$ and likewise it is impossible to embed $SU(3)\times SU(n)$ inside $E_8$ if $n>6$. We also note that the Coulomb branch of the $5d$ theories agrees with that expected from the $6d$ SCFTs. Before explaining this, we should first consider the $5d$ reduction in greater detail. Specifically, we noted that in $5d$ we see twisted affine groups. This suggests that the reduction must be done with a discrete symmetry twist which acts on these groups through their charge conjugation outer automorphism. There is a natural candidate for this discrete symmetry. Specifically, the $6d$ semi-gauge theory has a discrete symmetry acting on all the matter through charge conjugation, and it is natural to conjecture that it is present also in the $6d$ SCFT. In that case the $5d$ reduction should be done with a twist using that symmetry. 

We can now return to comparing the expected Coulomb branch dimension. As previously mentioned, when we compactify a $6d$ SCFT on a circle the Coulomb branch of the resulting $5d$ theory gets contributions from both the tensor branch and the vector multiplets on the circle, where the latter are also affected by the twist. The tensor branch of all the $6d$ SCFTs in the main line is three dimensional, so the Coulomb branch dimension of the expected $5d$ theory should be three plus the rank of the gauge symmetry invariant under the twist. Given a vector multiplet, we can parametrize the Coulomb branch by operators of the form $Tr(\phi^i)$, where $\phi$ is the scalar in the vector multiplet, which in our case comes from the component of the $6d$ vector in the circle direction. It is well known that for groups of type $SU(n)$, there are $n-1$ independent such operators, given by $i=2,3,4,...,n$. Out of these, the cases with $i$ even are invariant under the charge conjugation outer automorphism while cases with $i$ odd are not. When we compactify with a charge conjugation twist, only the cases that are invariant under the twist contribute to the Coulomb branch dimension. As a result, we expect one contribution from the $SU(3)$ vector multiplet, and $\frac{n}{2}$ or $\frac{n-1}{2}$ from the $SU(n)$ vector multiplet, depending on whether $n$ is even or odd.

Summing all the contributions we expect the Coulomb branch dimension of the resulting $5d$ theory to be $4+\frac{n}{2}$ for $n$ even or $4+\frac{n-1}{2}$ for $n$ odd. We can next compare it with the proposed $5d$ gauge theory. For a $Spin(n+8)$ gauge theory, the Coulomb branch dimension should be given by the rank of the gauge group, which is indeed $4+\frac{n}{2}$ for $n$ even or $4+\frac{n-1}{2}$ for $n$ odd.

Another check we can do is to consider Higgs branch flows. As we previously mentioned, the main line is connected via one such flow, where we give a vev to the gauge invariant made from a $Spin(n+8)$ vector hyper. We expect a similar flow pattern also for the $6d$ theories, and indeed such a flow pattern exists. Specifically, we can give a vev to a meson of the  $SU(n)$ group breaking it to $SU(n-1)$. This indeed initiates an analogous flow pattern. Interestingly, we can continue this line past the $n=1$ case. Recall that for $n=1$ we had a $Spin(9)$ $5d$ gauge theory with a single vector hyper and four spinor hypers. We can give a vev to the vector in the $Spin(9)$ theory leading to a $5d$ $Spin(8)$ gauge theory with four spinor hypers of both chiralities. We next consider the analogous flow in the conjectured $6d$ lift, which for this case is an $SU(3)$ gauging of the rank $2$ E-string theory. In this $6d$ SCFT, this Higgs branch flow should match giving a vev to the moment map operator associated with the $SU(2)$ global symmetry, which is the one reducing to the operator we gave a vev to in the $5d$ $Spin(9)$ gauge theory. This vev is known to break the rank $2$ E-string theory to two decoupled copies of the rank $1$ E-string theory. This suggests that the $5d$ $Spin(8)$ gauge theory with four spinor hypers of both chiralities lifts to a twisted compactification of a $6d$ SCFT with a tensor branch description as an $SU(3)$ gauging of two rank $1$ E-string theories, also known as the minimal $(E_6,E_6)$ conformal matter SCFT \cite{ZHTV}. 

This proposal can again be checked using similar methods as the previous case. First, we note that the $5d$ $Spin(8)$ gauge theory with four spinor hypers of both chiralities appears to be $6d$ lifting according to the criteria of \cite{Jefferson:2017ahm}. Also, it was found to have a perturbative plus 1-instanton spectrum of currents that is consistent with $E^{(2)}_6 \times E^{(2)}_6$ \cite{Zafrir:2015uaa}. Finally, we note that the Coulomb branch dimension matches that expected from the $6d$ SCFT. In light of this proposal, we can wonder whether an interesting enhancement of symmetry can be found in a $4d$ $\mathcal{N}=1$ $Spin(8)$ gauge theory with four spinor chirals of both chiralities, one that can then be also attributed to a similar $6d$ origin. We indeed find such a case which is further discussed in the appendix. 

One can also consider other flows, notably ones that in the $5d$ theory are associated with spinor vevs. For instance we can consider giving a spinor vev to one of the spinors in the $5d$ $Spin(8)$ gauge theory with four spinor hypers of both chiralities that we introduced previously. This just leads to a $5d$ $Spin(7)$ gauge theory with four spinor hypers and three vector hypers, as can be seen by using the triality automorphism of $Spin(8)$ to map the spinor vev to a vector one. This $5d$ gauge theory appeared already in \cite{Razamat:2018gbu}, where it was conjectured that it is $6d$ lifting and lifts to a twisted compactification of a $6d$ SCFT with tensor branch description as a gauging of a rank $1$ E-string SCFT by an $SU(3)$ gauge theory with six fundamental hypermultiplets. We can now see that this conjecture is compatible with our results. Specifically, the spinor vev we consider here should map to going on the Higgs branch of one of the rank $1$ E-string theories, where the choice of E-string theory mapping to the choice of the chirality of the spinor given a vev. This should then break the rank $1$ E-string theory to free hypers, which just become six fundamentals for the $SU(3)$ gauge group. We can similarly consider other spinor vevs, however, as this can be long and quite technical, we shall not discuss it in detail here. 

\subsubsection*{The two other $Spin(12)$ theories}

Finally, we want to consider the two other $Spin(12)$ theories. We can also find proposed $6d$ SCFT lifts for these cases. For the case with three half-hyper spinors in one chirality and one in the other, we propose the $6d$ lift is given again by a rank $1$ E-string theory gauged by a pure $SU(3)$ gauge theory, but also by a $G_2$ gauge theory with four fundamental hypermultiplets. For the case with spinor hypers of only one chirality we propose the $6d$ lift is given again by two rank $1$ E-string theories, both gauged by a pure $SU(3)$ gauge theory with an additional pure $SU(3)$ gauge theory gauging only one of them. When the appropriate tensors are added, both give anomaly free gauge theories, which should lift to $6d$ SCFTs. The resulting $6d$ SCFTs should have a global symmetry compatible with the $5d$ gauge theories. It is also straightforward to see that the resulting Coulomb branch dimensions match that expected from the $5d$ gauge theory. Finally both of these posses a Higgs branch flow leading to the main branch. In the former case, it is given by Higgsing the $G_2$ to $SU(3)$, while in the latter it is given by going on the Higgs branch of the rank $1$ E-string SCFT, gauged by only one $SU(3)$. This breaks the rank $1$ E-string SCFT to free hypermultiplets, some of which decouple, but those that do not become six fundamental hypermultiplets for the gauging $SU(3)$ group.    

\

\subsection{Aspects of the reduction}

Having formulated a conjecture for the $6d$ lift of the family of $Spin(n+8)$ gauge theories, we now want to use this conjecture to propose a general explanation for the origin of the $4d$ symmetry enhancement pattern \eqref{4dpatt8} found in the previous section. The idea is as follows. Consider a compactification of these theories on a torus with fluxes in the global symmetry. As we pointed out the global symmetry in this class of theories is $C_{E_{9-n}}\left(SU\left(3\right)\right)\times SU\left(2n\right)$, except for $n=2$. This global symmetry is expected to be inherited by the $4d$ theory, though part of it may be broken by the flux and other discrete options having to do with coupling to non-trivial flavor backgrounds (see for instance \cite{Bah:2017gph,Ohmori:2018ona}). Specifically, let us concentrate on flux, and consider turning on two independent fluxes, one in a $U(1)$ such that $SU\left(2n\right)$ is broken to $U(1)\times SU\left(n\right)^2$ and another in $C_{E_{9-n}}\left(SU\left(3\right)\right)$ such that it is broken to $U(1)\times C_{E_{9-n}}\left(SU\left(3\right)\times SU\left(2\right)\right)$. This should lead to a $4d$ theory with a global symmetry which is at least $U(1)^2\times C_{E_{9-n}}\left(SU\left(3\right)\times SU\left(2\right)\right) \times SU\left(n\right)^2$. We include here the possibility that one or both of these fluxes are zero and the symmetry enhances.

From the resulting $4d$ theories, we can generate additional theories by various deformations, like mass deformations, giving vevs to various operators or coupling to various additional free fields. In particular it is possible to use these deformations to get to a theory with $U(1)\times C_{E_{9-n}}\left(SU\left(3\right)\times SU\left(2\right)\right) \times SU\left(n\right)^2$ global symmetry, where one $U(1)$ is broken by the deformation. It is also possible to cause the breaking of the $6d$ global symmetry to this symmetry using the deformations instead of the fluxes. We note that the global symmetry $U(1)\times C_{E_{9-n}}\left(SU\left(3\right)\times SU\left(2\right)\right) \times SU\left(n\right)^2$ is the enhanced symmetry we found in the $4d$ $Spin(n+8)$ models, see Eq. \eqref{4dpatt8}. The only exceptional case is the $n=2$ one where, on one side the $6d$ global symmetry is enhanced, while on the other the $4d$ found enhanced symmetry is smaller. Presumably this should be accounted for by the deformation incidentally breaking more symmetry in this case. The matter content of these $Spin(n+8)$ models is, up to additional free fields, precisely an $\mathcal{N}=1$ $4d$ version of the matter content of the $5d$ theory we get via a twisted circle compactification. By our previous considerations then, it seems quite reasonable that these theories could arise from deformations of the $6d$ compactifications, and thus the enhancement of symmetry in them can be explained in this way.

We note that to get the $5d$ theories we need to perform a charge conjugation twist. This may explain why there is a rank enhancement. The general logic here is to build this theory by gluing two tubes each incorporating the charge conjugation twist. As each tube incorporates a twist, the punctures at the end should be described by boundary conditions for the twisted theory, that is the $5d$ $Spin(n+8)$ theories. The rank of the global symmetry visible in these theories is smaller than that of the $6d$ SCFT because of the twist. When we glue together two of these tubes we expect, on one hand, to get $4d$ $Spin(n+8)$ models with a matter content similar to the $4d$ theories considered here, though these in general will contain additional fields. On the other hand, as we are joining two tubes with a $\mathbb{Z}_2$ twist, the twist in the full surface vanishes and the symmetry should enhance back to the full $6d$ global symmetry compatible with the flux. This then leads to a large enhancement of symmetry in this theory and its descendants through various deformations preserving this symmetry.

This matching in behavior between the $4d$ enhanced symmetries and the global symmetries of the $6d$ SCFT lifts of the $\mathcal{N}=1$ $5d$ variants is the major indirect evidence for our proposed explanation. Unfortunately, as the symmetry of our $4d$ models is smaller than that of the direct compactification, some deformations must also be involved, though precisely which ones is unknown. Furthermore, from the models discussed here, it is possible to generate new ones with the same enhanced symmetry by flipping operators in complete representations of the enhanced groups. The choice of flipping we used in $4d$ is partially dictated by the desire to avoid having operators hitting the unitarity bound. However, this restriction is usually not obeyed in the torus compactifications of $6d$ SCFTs. The large number of possibilities regarding the uncertainties in the details of the $6d$ compactification makes it difficult to quantitatively check this proposal by, for example, anomaly computations. We do note that there are similar cases of enhancements, like the ones discussed in \cite{Razamat:2018gbu}, where some of the enhancements have been linked also quantitatively to a $6d$ reduction. In these cases the torus compactifications of the $6d$ SCFTs in question are better understood, allowing one to explicitly find the needed deformations. For now, we leave a more detailed study of this proposal for future work.

%% file: sections/appendix.tex
\section{Index calculation in detail}
\label{app}

In this appendix we consider a $Spin(8)$ model which can be regarded as the $n=0$ case of the sequence discussed in section \ref{fourdim}, and demonstrate the computation of its superconformal index in detail. Using this calculation, we will be able to examine some of its IR properties, and in particular its symmetry enhancement. 

The matter content is given in the following table,  
\begin{center}
	\begin{tabular}{|c||c|c|c|c|c|}
		\hline
		Field & $Spin\left(8\right)_{g}$ & $SU\left(4\right)_{s}$ & $SU\left(4\right)_{c}$ & $U\left(1\right)_{a}$ & $U\left(1\right)_{\hat{r}}$\\
		\hline 
		
		$S$ & $\boldsymbol{8}_{s}$ & $\boldsymbol{4}$ & $\boldsymbol{1}$ & 1 & $\frac{1}{4}$\\
		
		$C$ & $\boldsymbol{8}_{c}$ & $\boldsymbol{1}$ & $\boldsymbol{4}$ & -1 & $\frac{1}{4}$\\
		
		$F_{0}$ & $\boldsymbol{1}$ & $\boldsymbol{\overline{10}}$ & $\boldsymbol{1}$ & -2 & $\frac{3}{2}$\\
		
		$F_{1}$ & $\boldsymbol{1}$ & $\boldsymbol{1}$ & $\boldsymbol{\overline{10}}$ & 2 & $\frac{3}{2}$\\
		\hline
	\end{tabular}
	\label{table:Spin8}
\end{center}
The superpotential is given by 
\begin{equation}
W=F_{0}S^{2}+F_{1}C^{2},
\end{equation}
and using $a$ maximization \cite{Intriligator:2003jj} we find that $U\left(1\right)_{\hat{r}}$ is the superconformal R charge and that the conformal anomalies are $a=\frac{123}{64}$ and $c=\frac{163}{64}$. Moreover, all the gauge-invariant operators in this model are above the unitarity bound. 

\

Turning to the computation of the index, we first recall from subsection \ref{index} that chiral and vector multiplets contribute as follows,
\begin{table}[htbp]
	\begin{center}
		\begin{tabular}{|c|c|c|}
			\hline 
			Multiplet & Component & Contribution to the index\\
			\hline 
			\hline 
			\multirow{2}{*}{Chiral $\Phi$} & Scalar $\Phi$ & $+\left(pq\right)^{\frac{r}{2}}$\\
			\cline{2-3} 
			& Anti-spinor $\bar{\psi}_{\dot{+}}^{\Phi}$ & $-\left(pq\right)^{\frac{2-r}{2}}$\\
			\hline 
			\multirow{2}{*}{Vector $\mathcal{V}$} & Gaugino $\lambda_{+}$ , $\lambda_{-}$ & $-p$ , $-q$\\
			\cline{2-3} 
			& 
			$\partial_{-\dot{+}}\lambda_{+}+\partial_{+\dot{+}}\lambda_{-}=0$
			& $+2pq$\\
			\hline 
			\hline 
			\multicolumn{1}{c|}{} & Derivatives $\partial_{+\dot{+}}$ , $\partial_{-\dot{+}}$ & $+p$ , $+q$\\
			\cline{2-3} 
		\end{tabular}
		\caption{The various ingredients that have a nonvanishing contribution to the
			index. In this table, $r$ is the IR R charge of $\Phi$, and the
			characters corresponding to gauge and global symmetries are suppressed
			({\it e.g.} the gaugino is always in the adjoint of the gauge group, and
			thus its contribution will include the corresponding character). }
		\label{table:Cont}
	\end{center}
\end{table}

We begin with finding the coefficient at order $pq$ in the expansion of the index. To do that, we should first list all the gauge-invariant operators that contribute at this order, and find their representations under the global symmetries. Then, summing their contributions with a sign given by their fermion number $(-1)^{F}$ (recall the trace formula \eqref{indtrace}) would yield the desired result. Denoting the representations under the nonabelian groups in this model by $(\boldsymbol{R}_{Spin\left(8\right)_{g}},\boldsymbol{R}_{SU\left(4\right)_{s}},\boldsymbol{R}_{SU\left(4\right)_{c}})$, these operators are given in the table below. Note that only the form of the operators in terms of the various fields is written, and nonabelian indices are suppressed. In particular, when more than one representation appears for a given operator form, it should be taken as representations of different operators that have this same form. Moreover, the $U(1)_a$ charge of all the operators in the table below vanishes (the total contribution of other sectors of $U(1)_a$ charge at order $pq$ turns out to be zero). 
\begin{center}
	\begin{longtable}{|c|c|c|}
		\hline 
		$\mathrm{Operator}$ & $\left(-1\right)^{F}$ & $\mathrm{Representations}\,\,(R)$\tabularnewline
		\hline 
		\hline 
		$\lambda_{+}\lambda_{-}$ & + & $\left(\boldsymbol{1},\boldsymbol{1},\boldsymbol{1}\right)$\tabularnewline
		\hline 
		$\bar{\psi}_{\dot{+}}^{S}S$ & - & $\left(\boldsymbol{1},\boldsymbol{1},\boldsymbol{1}\right),\,\,\left(\boldsymbol{1},\boldsymbol{15},\boldsymbol{1}\right)$\tabularnewline
		\hline 
		$\bar{\psi}_{\dot{+}}^{C}C$ & - & $\left(\boldsymbol{1},\boldsymbol{1},\boldsymbol{1}\right),\,\,\left(\boldsymbol{1},\boldsymbol{1},\boldsymbol{15}\right)$\tabularnewline
		\hline 
		$\bar{\psi}_{\dot{+}}^{F_{0}}F_{0}$ & - & $\left(\boldsymbol{1},\boldsymbol{1},\boldsymbol{1}\right),\,\,\left(\boldsymbol{1},\boldsymbol{15},\boldsymbol{1}\right),\,\,\left(\boldsymbol{1},\boldsymbol{84},\boldsymbol{1}\right)$\tabularnewline
		\hline 
		$\bar{\psi}_{\dot{+}}^{F_{1}}F_{1}$ & - & $\left(\boldsymbol{1},\boldsymbol{1},\boldsymbol{1}\right),\,\,\left(\boldsymbol{1},\boldsymbol{1},\boldsymbol{15}\right),\,\,\left(\boldsymbol{1},\boldsymbol{1},\boldsymbol{84}\right)$\tabularnewline
		\hline 
		$F_{0}S^{2}$ & + & $\left(\boldsymbol{1},\boldsymbol{1},\boldsymbol{1}\right),\,\,\left(\boldsymbol{1},\boldsymbol{15},\boldsymbol{1}\right),\,\,\left(\boldsymbol{1},\boldsymbol{84},\boldsymbol{1}\right)$\tabularnewline
		\hline 
		$F_{1}C^{2}$ & + & $\left(\boldsymbol{1},\boldsymbol{1},\boldsymbol{1}\right),\,\,\left(\boldsymbol{1},\boldsymbol{1},\boldsymbol{15}\right),\,\,\left(\boldsymbol{1},\boldsymbol{1},\boldsymbol{84}\right)$\tabularnewline
		\hline 
		$S^{4}C^{4}$ & + & $\begin{array}{c}
		\left(\boldsymbol{1},\boldsymbol{1},\boldsymbol{1}\right),\,\,\left(\boldsymbol{1},\boldsymbol{1},\boldsymbol{35}\right),\,\,\left(\boldsymbol{1},\boldsymbol{1},\boldsymbol{20'}\right),\,\,\left(\boldsymbol{1},\boldsymbol{1},\boldsymbol{45}\right),\\
		2\left(\boldsymbol{1},\boldsymbol{15},\boldsymbol{15}\right),\,\,\left(\boldsymbol{1},\boldsymbol{45},\boldsymbol{15}\right),\,\,\left(\boldsymbol{1},\boldsymbol{15},\boldsymbol{45}\right),\,\,\left(\boldsymbol{1},\boldsymbol{45},\boldsymbol{45}\right),\\
		2\left(\boldsymbol{1},\boldsymbol{20'},\boldsymbol{20'}\right),\,\,\left(\boldsymbol{1},\boldsymbol{35},\boldsymbol{20'}\right),\,\,\left(\boldsymbol{1},\boldsymbol{20'},\boldsymbol{35}\right),\,\,\left(\boldsymbol{1},\boldsymbol{35},\boldsymbol{35}\right),\\
		\left(\boldsymbol{1},\boldsymbol{35},\boldsymbol{1}\right),\,\,\left(\boldsymbol{1},\boldsymbol{20'},\boldsymbol{1}\right),\,\,\left(\boldsymbol{1},\boldsymbol{45},\boldsymbol{1}\right)
		\end{array}$\tabularnewline
		\hline 
		$S^{2}C^{4}\bar{\psi}_{\dot{+}}^{F_{0}}$ & - & $\begin{array}{c}
		\left(\boldsymbol{1},\boldsymbol{15},\boldsymbol{15}\right),\,\,\left(\boldsymbol{1},\boldsymbol{45},\boldsymbol{15}\right),\,\,\left(\boldsymbol{1},\boldsymbol{15},\boldsymbol{45}\right),\,\,\left(\boldsymbol{1},\boldsymbol{45},\boldsymbol{45}\right),\\
		\left(\boldsymbol{1},\boldsymbol{20'},\boldsymbol{1}\right),\,\,\left(\boldsymbol{1},\boldsymbol{35},\boldsymbol{1}\right),\,\,\left(\boldsymbol{1},\boldsymbol{45},\boldsymbol{1}\right),\\
		\left(\boldsymbol{1},\boldsymbol{20'},\boldsymbol{20'}\right),\,\,\left(\boldsymbol{1},\boldsymbol{35},\boldsymbol{20'}\right),\,\,\left(\boldsymbol{1},\boldsymbol{45},\boldsymbol{20'}\right),\\
		\left(\boldsymbol{1},\boldsymbol{20'},\boldsymbol{35}\right),\,\,\left(\boldsymbol{1},\boldsymbol{35},\boldsymbol{35}\right),\,\,\left(\boldsymbol{1},\boldsymbol{45},\boldsymbol{35}\right)
		\end{array}$\tabularnewline
		\hline 
		$C^{4}\left(\bar{\psi}_{\dot{+}}^{F_{0}}\right)^{2}$ & + & $\begin{array}{c}
		\left(\boldsymbol{1},\boldsymbol{45},\boldsymbol{20'}\right),\,\,\left(\boldsymbol{1},\boldsymbol{45},\boldsymbol{35}\right)\end{array}$\tabularnewline
		\hline 
		$C^{2}\bar{\psi}_{\dot{+}}^{F_{1}}\left(\bar{\psi}_{\dot{+}}^{F_{0}}\right)^{2}$ & - & $\left(\boldsymbol{1},\boldsymbol{45},\boldsymbol{20'}\right),\,\,\left(\boldsymbol{1},\boldsymbol{45},\boldsymbol{35}\right),\,\,\left(\boldsymbol{1},\boldsymbol{45},\boldsymbol{45}\right)$\tabularnewline
		\hline 
		$\left(\bar{\psi}_{\dot{+}}^{F_{1}}\right)^{2}\left(\bar{\psi}_{\dot{+}}^{F_{0}}\right)^{2}$ & + & $\left(\boldsymbol{1},\boldsymbol{45},\boldsymbol{45}\right)$\tabularnewline
		\hline 
		$S^{2}C^{2}\bar{\psi}_{\dot{+}}^{F_{1}}\bar{\psi}_{\dot{+}}^{F_{0}}$ & + & $\begin{array}{c}
		\left(\boldsymbol{1},\boldsymbol{15},\boldsymbol{15}\right),\,\,\left(\boldsymbol{1},\boldsymbol{45},\boldsymbol{15}\right),\,\,\left(\boldsymbol{1},\boldsymbol{15},\boldsymbol{45}\right),\,\,\left(\boldsymbol{1},\boldsymbol{45},\boldsymbol{45}\right),\\
		\left(\boldsymbol{1},\boldsymbol{20'},\boldsymbol{20'}\right),\,\,\left(\boldsymbol{1},\boldsymbol{20'},\boldsymbol{35}\right),\,\,\left(\boldsymbol{1},\boldsymbol{20'},\boldsymbol{45}\right),\\
		\left(\boldsymbol{1},\boldsymbol{35},\boldsymbol{20'}\right),\,\,\left(\boldsymbol{1},\boldsymbol{35},\boldsymbol{35}\right),\,\,\left(\boldsymbol{1},\boldsymbol{35},\boldsymbol{45}\right),\\
		\left(\boldsymbol{1},\boldsymbol{45},\boldsymbol{20'}\right),\,\,\left(\boldsymbol{1},\boldsymbol{45},\boldsymbol{35}\right),\,\,\left(\boldsymbol{1},\boldsymbol{45},\boldsymbol{45}\right)
		\end{array}$\tabularnewline
		\hline 
		$S^{2}\left(\bar{\psi}_{\dot{+}}^{F_{1}}\right)^{2}\bar{\psi}_{\dot{+}}^{F_{0}}$ & - & $\left(\boldsymbol{1},\boldsymbol{20'},\boldsymbol{45}\right),\,\,\left(\boldsymbol{1},\boldsymbol{35},\boldsymbol{45}\right),\,\,\left(\boldsymbol{1},\boldsymbol{45},\boldsymbol{45}\right)$\tabularnewline
		\hline 
		$S^{4}C^{2}\bar{\psi}_{\dot{+}}^{F_{1}}$ & - & $\begin{array}{c}
		\left(\boldsymbol{1},\boldsymbol{15},\boldsymbol{15}\right),\,\,\left(\boldsymbol{1},\boldsymbol{45},\boldsymbol{15}\right),\,\,\left(\boldsymbol{1},\boldsymbol{15},\boldsymbol{45}\right),\,\,\left(\boldsymbol{1},\boldsymbol{45},\boldsymbol{45}\right),\\
		\left(\boldsymbol{1},\boldsymbol{1},\boldsymbol{20'}\right),\,\,\left(\boldsymbol{1},\boldsymbol{1},\boldsymbol{35}\right),\,\,\left(\boldsymbol{1},\boldsymbol{1},\boldsymbol{45}\right),\\
		\left(\boldsymbol{1},\boldsymbol{20'},\boldsymbol{20'}\right),\,\,\left(\boldsymbol{1},\boldsymbol{20'},\boldsymbol{35}\right),\,\,\left(\boldsymbol{1},\boldsymbol{20'},\boldsymbol{45}\right),\\
		\left(\boldsymbol{1},\boldsymbol{35},\boldsymbol{20'}\right),\,\,\left(\boldsymbol{1},\boldsymbol{35},\boldsymbol{35}\right),\,\,\left(\boldsymbol{1},\boldsymbol{35},\boldsymbol{45}\right)
		\end{array}$\tabularnewline
		\hline 
		$S^{4}\left(\bar{\psi}_{\dot{+}}^{F_{1}}\right)^{2}$ & + & $\left(\boldsymbol{1},\boldsymbol{20'},\boldsymbol{45}\right),\,\,\left(\boldsymbol{1},\boldsymbol{35},\boldsymbol{45}\right)$\tabularnewline
		\hline 
	\end{longtable}
\end{center}
Summing (with signs) the characters corresponding to the representations given in the table,\footnote{Note that we use the same notation for both characters and representations, and when we refer to the index we always mean the characters written in terms of the corresponding fugacities.} we obtain 
\begin{equation}
\sum R\left(-1\right)^{F}=\left(\boldsymbol{1},\boldsymbol{15},\boldsymbol{15}\right)+\left(\boldsymbol{1},\boldsymbol{20'},\boldsymbol{20'}\right)-\left(\boldsymbol{1},\boldsymbol{15},\boldsymbol{1}\right)-\left(\boldsymbol{1},\boldsymbol{1},\boldsymbol{15}\right).
\end{equation}
Now, as the coefficient at order $pq$ equals the contribution from the marginal operators minus that from the conserved currents (recall \eqref{indform}), we see that at a general point on the conformal manifold: 
\begin{equation}
\textrm{Marginals}=\left(\boldsymbol{1},\boldsymbol{15},\boldsymbol{15}\right)+\left(\boldsymbol{1},\boldsymbol{20'},\boldsymbol{20'}\right)+\left(\boldsymbol{1},\boldsymbol{1},\boldsymbol{1}\right),
\end{equation}
\begin{equation}
\textrm{Currents}=\left(\boldsymbol{1},\boldsymbol{15},\boldsymbol{1}\right)+\left(\boldsymbol{1},\boldsymbol{1},\boldsymbol{15}\right)+\left(\boldsymbol{1},\boldsymbol{1},\boldsymbol{1}\right),
\end{equation}
that is the symmetry is the same as in the UV, $SU\left(4\right)^{2}\times U\left(1\right)$. However, since we have a relevant operator of the form $S^{2}C^{2}$ that transforms in the $\left(\boldsymbol{1},\boldsymbol{6},\boldsymbol{6}\right)$ and have an IR R charge 1, we conjecture, as in the $Spin(12)$ theory of section \ref{fourdim}, that there is a point on the conformal manifold at which this operator gives rise to extra marginal operators of the form $(S^{2}C^{2})^{2}$ and we have: 
\begin{equation}
\label{margsp8}
\textrm{Marginals}=\left(\boldsymbol{1},\boldsymbol{1},\boldsymbol{20'}\right)+\left(\boldsymbol{1},\boldsymbol{20'},\boldsymbol{1}\right)+\left(\boldsymbol{1},\boldsymbol{15},\boldsymbol{15}\right)+\left(\boldsymbol{1},\boldsymbol{20'},\boldsymbol{20'}\right)+2\left(\boldsymbol{1},\boldsymbol{1},\boldsymbol{1}\right).
\end{equation}
Then, since $(\textrm{Marginals}-\textrm{Currents})$ is fixed along the conformal manifold, we get at this point: 
\begin{equation*}
\textrm{Currents}=\left(\boldsymbol{1},\boldsymbol{1},\boldsymbol{20'}\right)+\left(\boldsymbol{1},\boldsymbol{20'},\boldsymbol{1}\right)+\left(\boldsymbol{1},\boldsymbol{15},\boldsymbol{1}\right)+\left(\boldsymbol{1},\boldsymbol{1},\boldsymbol{15}\right)+2\left(\boldsymbol{1},\boldsymbol{1},\boldsymbol{1}\right)
\end{equation*}
\begin{equation}
=\boldsymbol{35}_{SU\left(6\right)_{1}}+\boldsymbol{35}_{SU\left(6\right)_{2}}+2,
\end{equation}
implying that the symmetry enhances to $SU\left(6\right)^{2}\times U\left(1\right)^{2}$. In terms of this symmetry, \eqref{margsp8} becomes:\footnote{Note that $\boldsymbol{15}_{SU\left(4\right)}$ can be identified as either $\boldsymbol{15}_{SU\left(6\right)}$ or $\boldsymbol{\overline{15}}_{SU\left(6\right)}$, and similarly $\boldsymbol{20'}_{SU\left(4\right)}+1$ can be identified as either $\boldsymbol{21}_{SU\left(6\right)}$ or $\boldsymbol{\overline{21}}_{SU\left(6\right)}$. We will write here the unbarred versions (the same applies also to $\boldsymbol{6}_{SU\left(4\right)}=\boldsymbol{6}_{SU\left(6\right)}$).}  
\begin{equation}
\textrm{Marginals}=\boldsymbol{21}_{SU\left(6\right)_{1}}\boldsymbol{21}_{SU\left(6\right)_{2}}+\boldsymbol{15}_{SU\left(6\right)_{1}}\boldsymbol{15}_{SU\left(6\right)_{2}}+1\,.
\end{equation}

\

We can now turn to other orders in the expansion of the index, and check that all the contributions indeed form representations of the larger symmetry. We will focus here on the lower orders, performing a similar computation to the one presented above. Starting with $(pq)^{1/2}$, the operators that contribute are as follows: 
\begin{center}
	\begin{tabular}{|c|c|c|}
		\hline 
		$\mathrm{Operator}$ & $\left(-1\right)^{F}$ & $\mathrm{Representations}\,\,(R)$\tabularnewline
		\hline 
		\hline 
		$S^{2}C^{2}$ & + & $\left(\boldsymbol{1},\boldsymbol{10},\boldsymbol{10}\right),\,\,\left(\boldsymbol{1},\boldsymbol{6},\boldsymbol{6}\right)$\tabularnewline
		\hline 
		$\bar{\psi}^{F_{0}}C^{2}$ & - & $\left(\boldsymbol{1},\boldsymbol{10},\boldsymbol{10}\right)$\tabularnewline
		\hline 
		$S^{2}\bar{\psi}^{F_{1}}$ & - & $\left(\boldsymbol{1},\boldsymbol{10},\boldsymbol{10}\right)$\tabularnewline
		\hline 
		$\bar{\psi}^{F_{0}}\bar{\psi}^{F_{1}}$ & + & $\left(\boldsymbol{1},\boldsymbol{10},\boldsymbol{10}\right)$\tabularnewline
		\hline 
	\end{tabular}
\end{center}
Summing (with signs) these representations, we obtain 
\begin{equation}
\sum R\left(-1\right)^{F}=\left(\boldsymbol{1},\boldsymbol{6},\boldsymbol{6}\right)=\boldsymbol{6}_{SU\left(6\right)_{1}}\boldsymbol{6}_{SU\left(6\right)_{2}}\,.
\end{equation}

Next, at order $(pq)^{3/4}$ the operators divide into two groups, the first one has $U(1)_a$ charge 2 and the other one -2. The operators of the first group are
\begin{center}
	\begin{tabular}{|c|c|c|}
		\hline 
		$\mathrm{Operator}$ & $\left(-1\right)^{F}$ & $\mathrm{Representations}\,\,(R)$\tabularnewline
		\hline 
		\hline 
		$F_{1}$ & + & $\left(\boldsymbol{1},\boldsymbol{1},\boldsymbol{\overline{10}}\right)$\tabularnewline
		\hline 
		$S^{4}C^{2}$ & + & $\begin{array}{c}
		\left(\boldsymbol{1},\boldsymbol{20'},\boldsymbol{10}\right),\,\,\left(\boldsymbol{1},\boldsymbol{35},\boldsymbol{10}\right),\,\,\left(\boldsymbol{1},\boldsymbol{15},\boldsymbol{6}\right),\,\,\left(\boldsymbol{1},\boldsymbol{45},\boldsymbol{6}\right),\,\,\left(\boldsymbol{1},\boldsymbol{1},\boldsymbol{10}\right)\end{array}$\tabularnewline
		\hline 
		$S^{2}C^{2}\bar{\psi}^{F_{0}}$ & - & $\left(\boldsymbol{1},\boldsymbol{20'},\boldsymbol{10}\right),\,\,\left(\boldsymbol{1},\boldsymbol{35},\boldsymbol{10}\right),\,\,\left(\boldsymbol{1},\boldsymbol{45},\boldsymbol{10}\right),\,\,\left(\boldsymbol{1},\boldsymbol{15},\boldsymbol{6}\right),\,\,\left(\boldsymbol{1},\boldsymbol{45},\boldsymbol{6}\right)$\tabularnewline
		\hline 
		$C^{2}\left(\bar{\psi}^{F_{0}}\right)^{2}$ & + & $\left(\boldsymbol{1},\boldsymbol{45},\boldsymbol{10}\right)$\tabularnewline
		\hline 
		$\bar{\psi}^{F_{1}}\left(\bar{\psi}^{F_{0}}\right)^{2}$ & - & $\left(\boldsymbol{1},\boldsymbol{45},\boldsymbol{10}\right)$\tabularnewline
		\hline 
		$S^{2}\bar{\psi}^{F_{1}}\bar{\psi}^{F_{0}}$ & + & $\left(\boldsymbol{1},\boldsymbol{20'},\boldsymbol{10}\right),\,\,\left(\boldsymbol{1},\boldsymbol{35},\boldsymbol{10}\right),\,\,\left(\boldsymbol{1},\boldsymbol{45},\boldsymbol{10}\right)$\tabularnewline
		\hline 
		$S^{4}\bar{\psi}^{F_{1}}$ & - & $\left(\boldsymbol{1},\boldsymbol{20'},\boldsymbol{10}\right),\,\,\left(\boldsymbol{1},\boldsymbol{35},\boldsymbol{10}\right)$\tabularnewline
		\hline 
	\end{tabular}
\end{center}
and the sum of representations is 
\begin{equation}
\sum R\left(-1\right)^{F}=\begin{array}{c}
\left(\boldsymbol{1},\boldsymbol{1},\boldsymbol{10}\right)\end{array}+\left(\boldsymbol{1},\boldsymbol{1},\boldsymbol{\overline{10}}\right)=\boldsymbol{20}_{SU\left(6\right)_{2}}\,.
\end{equation}
The second group of operators is 
\begin{center}
	\begin{tabular}{|c|c|c|}
		\hline 
		$\mathrm{Operator}$ & $\left(-1\right)^{F}$ & $\mathrm{Representations}\,\,(R)$\tabularnewline
		\hline 
		\hline 
		$F_{0}$ & + & $\left(\boldsymbol{1},\boldsymbol{\overline{10}},\boldsymbol{1}\right)$\tabularnewline
		\hline 
		$S^{2}C^{4}$ & + & $\begin{array}{c}
		\left(\boldsymbol{1},\boldsymbol{10},\boldsymbol{20'}\right),\,\,\left(\boldsymbol{1},\boldsymbol{10},\boldsymbol{35}\right),\,\,\left(\boldsymbol{1},\boldsymbol{6},\boldsymbol{15}\right),\,\,\left(\boldsymbol{1},\boldsymbol{6},\boldsymbol{45}\right),\,\,\left(\boldsymbol{1},\boldsymbol{10},\boldsymbol{1}\right)\end{array}$\tabularnewline
		\hline 
		$S^{2}C^{2}\bar{\psi}^{F_{1}}$ & - & $\left(\boldsymbol{1},\boldsymbol{10},\boldsymbol{20'}\right),\,\,\left(\boldsymbol{1},\boldsymbol{10},\boldsymbol{35}\right),\,\,\left(\boldsymbol{1},\boldsymbol{10},\boldsymbol{45}\right),\,\,\left(\boldsymbol{1},\boldsymbol{6},\boldsymbol{15}\right),\,\,\left(\boldsymbol{1},\boldsymbol{6},\boldsymbol{45}\right)$\tabularnewline
		\hline 
		$S^{2}\left(\bar{\psi}^{F_{1}}\right)^{2}$ & + & $\left(\boldsymbol{1},\boldsymbol{10},\boldsymbol{45}\right)$\tabularnewline
		\hline 
		$\bar{\psi}^{F_{0}}\left(\bar{\psi}^{F_{1}}\right)^{2}$ & - & $\left(\boldsymbol{1},\boldsymbol{10},\boldsymbol{45}\right)$\tabularnewline
		\hline 
		$C^{2}\bar{\psi}^{F_{0}}\bar{\psi}^{F_{1}}$ & + & $\left(\boldsymbol{1},\boldsymbol{10},\boldsymbol{20'}\right),\,\,\left(\boldsymbol{1},\boldsymbol{10},\boldsymbol{35}\right),\,\,\left(\boldsymbol{1},\boldsymbol{10},\boldsymbol{45}\right)$\tabularnewline
		\hline 
		$C^{4}\bar{\psi}^{F_{0}}$ & - & $\left(\boldsymbol{1},\boldsymbol{10},\boldsymbol{20'}\right),\,\,\left(\boldsymbol{1},\boldsymbol{10},\boldsymbol{35}\right)$\tabularnewline
		\hline 
	\end{tabular}
\end{center}
and the sum is 
\begin{equation}
\sum R\left(-1\right)^{F}=\begin{array}{c}
\left(\boldsymbol{1},\boldsymbol{10},\boldsymbol{1}\right)\end{array}+\left(\boldsymbol{1},\boldsymbol{\overline{10}},\boldsymbol{1}\right)=\boldsymbol{20}_{SU\left(6\right)_{1}}\,.
\end{equation}

\

Now we can finally collect the results and write the expansion of the index up to order $pq$ (or up to any other order if we continue this calculation in the same way). Denoting the $U(1)_a$ fugacity by $a$ and using the representations of the larger symmetry $SU\left(6\right)_{1}\times SU\left(6\right)_{2}\times U\left(1\right)^{2}$, we obtain 
\begin{equation*}
\mathcal{I}=1+\boldsymbol{6}_{SU\left(6\right)_{1}}\boldsymbol{6}_{SU\left(6\right)_{2}}\left(pq\right)^{\frac{1}{2}}+\left(\boldsymbol{20}_{SU\left(6\right)_{1}}a^{-2}+\boldsymbol{20}_{SU\left(6\right)_{2}}a^{2}\right)\left(pq\right)^{\frac{3}{4}}+
\end{equation*}
\begin{equation}
+\left(\boldsymbol{15}_{SU\left(6\right)_{1}}\boldsymbol{15}_{SU\left(6\right)_{2}}+\boldsymbol{21}_{SU\left(6\right)_{1}}\boldsymbol{21}_{SU\left(6\right)_{2}}-\boldsymbol{35}_{SU\left(6\right)_{1}}-\boldsymbol{35}_{SU\left(6\right)_{2}}-1\right)pq+\ldots ,
\end{equation}
where we again stress that this is up to a charge conjugation on each representation as we cannot distinguish between an $SU(6)$ representation and its conjugate based only on the characters of its $SO(6)$ subgroup.

Finally, we note that this enhancement may also be explained by a $6d$ origin. Specifically, we noted in section \ref{sixdim} that the $5d$ $\mathcal{N}=1$ version of this theory appears to lift to a twisted compactification of a $6d$ SCFT with $E_6 \times E_6$ global symmetry. It is then tempting to suspect that this $4d$ theory might be generated by  the compactification of the $6d$ SCFT on a torus with fluxes, or at least by some deformation of it. In that case, the $E_6 \times E_6$ global symmetry can be broken to $SU(6)\times SU(6)$, plus abelian factors, by the deformation or flux, which will then explain the speculated enhancement.